\documentclass[12pt,preprint]{aastex}
\usepackage{color}
\newcommand{\app}{$\sim$\ }
\newcommand{\gapp}{$\gtrsim$\ }
\newcommand{\lapp}{$\lesssim$\ }

\newcommand{\macc}{\dot{M}_{acc}}
\newcommand{\mpy}{{\rm M}_{\odot} {\rm yr}^{-1}}
\newcommand{\ms}{${\rm M}_{\odot}$}
\newcommand{\ls}{${\rm L}_{\odot}$}
\newcommand{\mst}{${\rm M}_{*}$\ }
\newcommand{\pow}[2]{$ #1 \times 10^{#2}$}
\newcommand{\beq}{\begin{equation}}
\newcommand{\eeq}{\end{equation}}

\begin{document}
\title{Time-Evolution of Viscous Circumstellar Disks due to Photoevaporation 
 by FUV, EUV and X-ray Radiation from the Central Star} 
\author{U.~Gorti\altaffilmark{1,2} }
\author{C.~P.~Dullemond\altaffilmark{3}}
\author{D.~Hollenbach\altaffilmark{2}} 
\altaffiltext{1}{NASA Ames Research Center, Moffett Field, CA}
\altaffiltext{2}{SETI Institute, Mountain View, CA}
\altaffiltext{3}{Max Planck Institute for Astronomy, Heidelberg, Germany}   
\begin{abstract}
We present the time evolution of viscously accreting circumstellar disks as they are irradiated by ultraviolet and X-ray photons from a low-mass central star.  Our model is a hybrid of a 1D time-dependent viscous disk model coupled to a 1+1D disk vertical structure model used for calculating the disk structure and photoevaporation rates.    We find that disks of initial mass $0.1$\ms\   around \app 1\ms\ stars survive for \app \pow{4}{6} years, assuming a viscosity parameter $\alpha=0.01$, a time-dependent FUV luminosity $L_{FUV}\sim 10^{-2}-10^{-3}$ \ls\ and   with X-ray and EUV luminosities $L_X \sim L_{EUV}  \sim 10^{-3}$L$_{\odot}$.  We find   that FUV/X-ray-induced photoevaporation and viscous accretion are
 both important in  depleting disk mass. Photoevaporation rates are most significant at \app 1-10 AU and at \gapp 30 AU. Viscosity spreads the disk which causes mass loss by accretion onto the central star and feeds mass loss by photoevaporation in the outer disk. 
  We find that FUV photons can create gaps in  the inner, planet-forming regions of the disk ($\sim 1-10$ AU) at relatively early epochs in disk evolution while disk masses are still substantial.  EUV and X-ray photons are also capable of driving gaps, but EUV can only do so at late, low accretion-rate  epochs after the disk mass has already declined substantially.  Disks around stars with predominantly soft X-ray fields experience enhanced photoevaporative mass loss. We follow disk evolution around stars of different masses, and find that disk survival time is relatively  independent of mass for stars with \mst \lapp 3\ms; for \mst \gapp 3\ms\ the disks  are short-lived($ \sim 10^5$ years).   
  
\end{abstract}
\keywords{accretion, accretion disks --- stars: formation --- planetary systems:protoplanetary disks --- stars: pre-main-sequence --- ultraviolet:stars --- X-rays:stars}

\section{Introduction}
Circumstellar disks are a natural outcome of the star formation process, but are dispersed on timescales short (\lapp \pow{5}{6} years)  compared to the lifetimes of their central stars (e.g., Haisch et al. 2001, Hillenbrand 2005).  Disks undergo significant evolution in their structure and composition as they age, and are also believed to be the birthplaces of planetary systems. Understandably, disk evolution and planet formation are inextricably linked processes. Disk lifetimes limit planet formation timescales and studies of disks provide invaluable clues to the requisite conditions essential to planetary system formation. The manner in which disks disperse and the timescale of the removal process is vital to understanding disk evolution and planet formation. 

Photoevaporation\footnote{Photoevaporation is the process by which energetic (UV or X-ray) photons  from the central star or a nearby massive star heat the surface of the disk and cause thermal-pressure-driven hydrodynamic mass outflows  from the outer regions of the disk, where the escape speeds are lower or  comparable to the thermal speed.} of  protoplanetary disks is believed to be the most viable and, in fact, dominant mechanism for the dispersal of the outer disk where most of the mass resides (e.g., Hollenbach et al. 2000, Dullemond et al. 2007).  Earlier research focused on  photoevaporation caused by massive O and B type stars due to their significant ionizing radiation fields (Extreme Ultraviolet or EUV, $h\nu > 13.6$ eV), resulting in heating and mass loss of disk gas from both their own circumstellar disks (e.g., Hollenbach et al. 1994, Richling \& Yorke 1997, 1998, 2000) and those around nearby low-mass stars (e.g.,  Johnstone et al. 1998, Stoerzer and Hollenbach 1999, Johnstone et al. 2004).  Subsequent observations of disks around low mass stars near the bright Trapezium
O star $\Theta ^1$C have  confirmed the occurence of the process (e.g., Bally et al. 2000, Goto et al. 2006, Cieza et al. 2008) and observed mass loss rates correspond very well to the theoretically predicted rates (Rigliaco et al. 2009).  The success of this theory coupled with the observationally inferred rapid dispersal timescales 
of disks even in low-mass star forming regions led Clarke et al. (2001) to investigate EUV photoevaporation due to the radiation field of a central low mass host star itself. They combined photoevaporation with disk viscosity to find that EUV forms gaps in the inner disk 
(the so-called ``EUV switch'') and that disk lifetimes are of the order of  $10^7$ years. Alexander at al. (2006a,b; hereafter together referred to as ACP06) refined the theory to include direct irradiation of the disk inner rim once a gap is formed.   
The main idea of this work was that photoevaporation from the inner rim led to a relatively rapid dispersal of the outer disk once the gap forms. 
However, EUV photoevaporation  can  explain observed disk lifetimes only if the EUV fields are unrealistically high or if the initial disk mass is unusually low. EUV photoevaporation 
timescales are found to be \gapp $10^7$ years for realistic EUV photon luminosities ($\phi_{EUV}\sim10^{41-42}$ s$^{-1}$) and initial disk masses ($\sim 0.1 {\rm M}_*$), much longer than the observationally inferred disk lifetimes of $\sim$ few Myrs (e.g., Haisch et al. 2001, Hillenbrand 2005).   In a recent paper, Gorti and Hollenbach (2009; hereafter GH09)  show that FUV (Far Ultraviolet, 6eV$<h\nu<$13.6eV) and X-ray radiation from the central star early in its evolution causes significant photoevaporation from the disk and that they dominate the mass loss rates over  EUV-induced photoevaporation for most of the outer disk ($\gtrsim 2$ AU) where the disk mass reservoir lies.  Disk lifetimes are estimated to be $\sim$ few $10^6$ yrs for typical disk and stellar parameters.
The effect of X-rays alone on photoevaporation was examined first by Alexander et al. (2004) and more recently by Ercolano et al. (2008, 2009).  Ercolano et al. (2009, hereafter ECD09) find significant mass loss due to photoevaporation by soft (\app 0.1-0.3 keV) X-ray photons ($ \sim 10^{-9}\ \mpy$), if they are present and luminous. 

GH09  calculate the rate of photoevaporation of a circumstellar disk by energetic radiation (EUV, FUV and X-rays) from its central star, using detailed but {\em static} disk structure models (Gorti \& Hollenbach 2008, hereafter GH08).  We found in our static analysis that contrary to the EUV photoevaporation scenario, which creates a gap at about  $\sim 1-2$ AU for a 1\ms\ star and then erodes the outer disk from inside out, FUV photoevaporation predominantly removes less bound gas from the outer disk where most of the mass is typically located. Relatively hard($\sim$ 1keV) X-rays affect the disk  only indirectly by ionizing the gas, and  enhancing FUV-induced grain photoelectric heating (see also ECD09).   The creation of gaps in the inner disk by FUV and X-ray photons was suggested, but not conclusive in our static disk models. Photoevaporation timescales for disks around 1\ms \ stars were estimated to be a  $\sim {\rm few\ }10^6$ years, after the onset of disk irradiation by FUV and X-rays.  These results were  obtained by qualitative estimates of the mass loss rates at a representative disk ``epoch'' and were not time-dependent calculations, and did not include viscous evolution. 

The FUV luminosity  of the young star is, however, time-dependent because a large fraction arises  from the time-variable mass accretion rate  onto the central star. The FUV radiation incident on the disk  is high in the earliest stages of low mass star formation when accretion rates are high, and declines steadily to plateau at the chromospheric FUV level at late stages when accretion ceases. Accretion is usually accompanied by mass loss in the form of a disk wind (e.g., Shang et al. 2007) which can contain enough column density (also varying with time) to attenuate the stellar UV and X-ray photons when mass loss rates are high in the early evolution of the protostar/disk (Hollenbach \& Gorti 2009).  Determining the FUV/X-ray  photoevaporative disk dispersal timescales thus necessitates time-dependent calculations that  follow the evolution of the disk surface density as the incident radiation field changes with time. In addition, viscous spreading continuously affects the surface density distribution of the photoevaporating disk and in turn the resulting photoevaporation rates. 

This is the first in a series of papers where we study the photoevaporation of viscously evolving disks subject to EUV, FUV and X-ray radiation from the central star.  Our ultimate aim is to follow the evolution of the disk surface density, the hydrodynamics of the escaping flow, the evolution in disk dust properties such as opacity, and obtain spectral energy distributions of disk at various epochs for comparison with observations.  In this paper, as an initial step towards this goal,   we solve for the evolution of the surface density distribution of the disk in a simple 1-D radial model for viscous transport, coupled with a  1+1D  gas and dust radiative transfer model for determining the photoevaporation rates.   We do not explicitly treat the hydrodynamical flow, but use analytic approximations to estimate the mass flux from each radial annulus. The dust is assumed well-mixed with the gas with fixed opacity per hydrogen nucleus in time and space. 

The paper is organized as follows. In \S 2 we describe our model and assumptions involved. We then present and discuss our findings in \S 3. We end with a summary (\S 4)  and indicate  the future direction of our work.

\section{Model description}
We consider the time evolution of the radial surface density distribution of the disk in a simple 1-dimensional model, as first considered by Lynden-Bell \& Pringle (1974; hereafter LBP74).  The disk is subject to viscous evolution and is irradiated at the surface by stellar radiation driving photoevaporation as it evolves.   We use the usual simple $\alpha$-parametrization for disk viscosity $\nu$ (Shakura \& Sunyaev 1976).  The evolution of surface density distribution in the disk is then determined by the diffusion equation (LBP74) with a sink term   $\dot{\Sigma}_{pe}$ 
\begin{equation}
{{\partial \Sigma}\over{\partial t}} = {{3}\over{r} }{{\partial}\over{\partial r}} \left(
\sqrt r {{\partial}\over{\partial r}} \left( \nu \Sigma \sqrt r \right) \right) - \dot{\Sigma}_{pe} (r,t),
\label{sdot}
\end{equation}
where $r$ is the radial co-ordinate, $\Sigma$ is the surface density and $\dot{\Sigma}_{pe}$ is the instantaneous photoevaporation rate at that radius (from both sides of the disk) due to EUV, FUV and X-ray photoevaporation. 

Equation~\ref{sdot} is solved numerically, using an implicit integration  technique for numerical stability at large time steps. The viscosity  coefficient $\nu\equiv \alpha c_s^2/\Omega_K$, where $c_s$ is the  isothermal sound speed computed from the temperature of the disk at the midplane and $\Omega_K$  is the angular frequency for a Keplerian disk.  The  midplane temperature is calculated using a simple two-layer  approximation similar to that described in Chiang \& Goldreich (1997)  but with some adaptations. The Chiang \& Goldreich model depends  on the determination of an incidence angle of the stellar radiation  heating the disk, which could potentially become negative in the outer regions of the disk. 
This may then result in self-shadowing and disk solutions would require full 2-D  radiative transfer. Since 2-D radiative transfer calculations are  too demanding in terms of computing time for a disk evolution model, we choose a fixed  incidence angle that corresponds to typical values found in self-consistent 2-D disk models. The errors in the midplane temperature of the  disk are likely to be small enough for our model to keep its validity.  In addition to irradiational heating we also include viscous heating  
near the midplane, using the standard Shakura \& Sunyaev (1973) method  (see also Hubeny 1990). Such viscous heating will affect the disk  midplane temperature mostly in the inner disk regions, and when  accretion rates are still appreciable. In the outer regions of the  disk irradiation dominates entirely. Our model follows mostly the  
model of Hueso \& Guillot (2005), and was used by us in two earlier  publications (Dullemond, Natta \& Testi 2006 and Dullemond, Apai \&  Walch 2006). A more complete model description will be given in a near-future publication (Birnstiel, Dullemond \& Brauer in prep.).

We then calculate the instantaneous photoevaporation rate, $\dot{\Sigma}_{pe}$. 
In the case of FUV and X-ray heating, gas temperatures in the upper layers of the disk (extinction to the star A$_{\rm V} \sim 10-10^{-2}$) can range from $\sim$ a few 100K to $\sim$ a few 1000K, unlike the EUV case  where the gas temperature at the very surface, (A$_{\rm V} \lesssim 10^{-2}$) is nearly constant at $\sim 10^4$K. 
The vertical structure of the disk can therefore be regarded as consisting of three regions: from bottom to top they are: (i) a relatively cool midplane region where gas and dust temperatures are very nearly equal, (ii) a warmer surface layer where the dust solid-state spectral features are formed and where FUV/X-ray heating dominates, causing the gas temperature to deviate from  the dust temperature, and (iii) finally a very tenuous ionized and hot EUV-heated layer.
As described in detail in GH09, the determination of $\dot{\Sigma}_{pe}$  therefore involves solving for the disk vertical structure and obtaining the density $n$ and gas temperature $T_{gas}$ as a function of spatial location $(r,z)$,  and in this case  also as a function of time. Note that we solve for the {\em gas} temperature to determine the vertical structure in a self-consistent manner, and do not assume that the gas and dust temperatures are equal in order to set the vertical density profile, as is customary. We show in GH09 that the gas and dust temperatures can substantially deviate at the disk surface and significantly affect the determination of mass loss rates due to photoevaporation. $\dot{\Sigma}_{pe}$ at every radius is determined as follows.  At each timestep, we solve for the disk structure and calculate the density and temperature structure of gas (and dust). We then use the analytical approximations of Adams et al. (2004) to calculate the potential mass loss rate at every spatial location $(r,z)$ 
(also see GH09). Therefore,
\begin{equation}
\label{sigmadot2}
\dot{\Sigma}_{pe}\sim \mu n_s c_s(r,z) \left( {r_s}\over{r}\right)^2 
\end{equation}
where $n_s$ the density at the sonic radius $r_s$ and $c_s$ is the sound speed. 
$\mu$ is the mean mass per particle in the gas. As in GH09, for isothermal flow  from the launch point $(r,z)$ with density $n(r,z)$ to $r_s$, and for $r_s \gg r$, we obtain \begin{equation}
\label{dsonic}
n_s = n(r,z) \exp \left( -  {{r_g}\over{2r}}(1-{{r}\over{r_s}})^2\right)
\end{equation}
where $r_g =GM_*/c_s^2$ is the gravitational radius and $r_s$ is given by 
\begin{equation}
\label{rsonic}
r_s = {{r_g}\over{4}}\left(1+\left(1-{{8r}\over{r_g}}\right)^{1/2}\right)
\end{equation}
for $r<r_g/8$.   For $r>r_g/2$, the flow rapidly goes through a sonic point near the base and $r_s=r$.  For $r_g/8 < r < r_g/2$ we linearly extrapolate between $r_s = r_g/4$ at $r=r_g/8$ and $r_s=r_g/2$ at $r=r_g/2$. For FUV and X-ray heated regions, we use our disk structure models to determine the density and temperature as described earlier.  For EUV-heated flow we follow the analysis of Hollenbach et al. (1994) and the density at the base of the ionized layer where the flow originates is given by 
\begin{equation}
n_{II} \simeq 0.3 \left( { {3 \phi_{EUV}} \over{4 \pi \alpha_r r_{g,II}^3}}\right)^{1/2} \left( r/r_{g,II} \right)^{-p}
\label{nhii}
\end{equation}
where $\alpha_r =2.53\times 10^{-13} {\rm cm}^{3} {\rm s}^{-1}$ is the case B recombination coefficient of hydrogen, $r_{g,II} = GM_*/c_{s,II}^2$  is the gravitational radius for ionized gas,  $\phi_{EUV}$ s$^{-1}$ is the EUV photon luminosity of the star  and the exponent $p$  is equal to 1.5 for $r<r_{g,II}$ and
2.5 for $r>r_{g,II}$ (Hollenbach et al. 1994).  
Using Eqs.~(\ref{sigmadot2}-\ref{nhii}) and the density and temperature distribution from our disk models,  the  mass loss rate at each radius (from both surfaces of the disk) is determined by the $z$ layer with the highest  $\dot{\Sigma}_{pe}(r,t)$, which could be heated by FUV, EUV, X-rays or any of the other heating processes (\S 2.4). 
 This value is used in Eq.~\ref{sdot} to find the rate of change of surface density. We repeat this process to follow the evolution of the surface density distribution in time. 
 
Our models have two principal inputs: (i) stellar properties, viz., radiation field and mass and (ii) disk properties, characterized mainly by the initial disk mass, viscosity parameter $\alpha$ that sets the accretion rate onto the star, wind mass loss rate that is correlated with the accretion rate, gas phase abundances of elements  and the dust grain opacity per H nucleus.  
 
 \subsection{Stellar Radiation Field}
A young star is chromospherically active and emits substantial high-energy radiation in the X-ray and UV bands (e.g., Feigelson et al. 2003); in addition,  disk material accreting onto the stellar surface creates hotspots that generate copious amounts of energetic photons.  However, as  accretion is accompanied by a protostellar wind (likely generated magnetically in the inner disk,  e.g. Shu et al. 2000) with rates $\sim 0.1 \dot{M}_{acc}$ (e.g., White \& Hillenbrand 2004), there is significant opacity in the wind at early epochs to absorb the high energy radiation before it is incident on the outer disk. Accretion rates decrease with time and as the column density in the associated wind correspondingly decreases, FUV, X-ray and EUV photons begin to penetrate the wind and irradiate the disk. For typical wind parameters,  the accretion rates have to be 
$\lesssim 10^{-7}\ \mpy$ for disk irradiation by  FUV and hard X-ray ($E_X\gtrsim1$keV) photons and $\lesssim 10^{-8}\  \mpy$ for EUV and soft X-ray ($E_X \sim 0.1-0.5 $keV)  photons respectively (Hollenbach \& Gorti 2009). 

\paragraph{FUV} FUV luminosities of stars are well-studied observationally (e.g., IUE, Valenti et al. 2003; FUSE, Bergin et al. 2003, Herczeg et al. 2004) and theoretically believed to arise due to  both activity in the chromosphere and accretion hotspots on the surface of the star.  We consider both components  for our FUV field. In early stages of evolution, the accretion component dominates.   An important addition to our model is the calculation of the  time-dependent accretion luminosity from the accretion rate onto the star and hence, a consistent determination of the FUV luminosity.    We calculate accretion-generated FUV luminosities  from the mass accretion rate onto the star (at our innermost radial gridpoint) as determined by our disk evolution equation (Eq.~\ref{sdot}). We assume that the accretion luminosity can be approximated as a black body of temperature 9000K (Calvet \& Gullbring 1998; also see GH09). We then estimate the fraction of this luminosity in the FUV band (4\%) to obtain  
\begin{equation}
L_{FUV}^{acc} \approx 0.04\left({{ 0.8 G M_* \dot{M}_{acc}}\over{R_*}}\right) = 10^{-2} \left({{M_*}\over{M_{\odot}}} \right) \left({R_*}\over{R_{\odot}}\right)^{-1}
\left( {{ \dot{M}_{acc} }\over{10^{-8} \mpy}} \right) L_{\odot}.
 \end{equation} 
  The chromospheric component to the FUV flux has an FUV luminosity $L_{FUV}^{Chr.}$ given by  $\log(L_{FUV}^{Chr.}/{\rm L}_*)=-3.3$ (from the data for non-accreting, weak-line T Tauri stars in Valenti et al. 2003), similar to the observed scaling for X-ray luminosity. For massive stars with high stellar effective temperatures, we also add the stellar photospheric contribution to the FUV, $L_{FUV}^{*}$. The UV component of early-type stars is quite different from that estimated by a simple black body spectrum, and we therefore use data from Parravano et al.(2003) for the FUV and EUV fluxes from massive stars.  Therefore the total FUV luminosity is the sum of all three components and is time-dependent, 
  \begin{equation}
  L_{FUV}(t) = L_{FUV}^{acc}(t) + L_{FUV}^{Chr.} + L_{FUV}^*. 
  \end{equation}  
\paragraph{X-rays}
Measured X-ray fluxes from young stars are high (e.g. ROSAT, Chandra, and XMM-Newton data). X-rays are observed to be strong at all protostellar evolutionary stages, typically $\sim3$ orders of magnitude higher than  for main-sequence stars. X-ray luminosities of stars are a function of  stellar mass ($L_X \sim 2.3 \times 10^{30} {\rm  (M_*/M_{\odot})^{1.44} erg \ s}^{-1} $ for ${\rm M_* \lesssim 3 M_{\odot}}$ and $L_X \sim 10^{-6} L_*$ for ${\rm M_* \gtrsim 3 M_{\odot}}$;  Flaccomio et al. 2003, Preibisch et al. 2005), although there are large variations and short-period fluctuations especially for the youngest stars. We neglect X-ray variability and the occurrence of flares in the present analysis since the photoevaporation process is primarily sensitive to the mean field, and consider the X-ray luminosity for a star of given mass to be represented by the above relations. We  adopt a standard X-ray spectrum  that peaks at 2keV as in GH09, with $L_X(E) \propto E$ for $0.1$keV$<E <2$keV and $L_X(E) \propto E^{-1.75}$ for 2keV$<E<10$keV.  

Large X-ray surveys such as the Chandra Orion Ultradeep Project (COUP, Getman et al. 2005) and the XMM-Newton Extended Survey of the Taurus Molecular Cloud  (XEST, G\"udel et al. 2008)  reveal systematic differences in the X-ray properties of accreting and non-accreting T Tauri stars. Accreting T Tauri stars (CTTS) show some evidence of a soft X-ray excess ($\sim 0.2$ keV; e.g., Kastner et al. 2002, Stelzer \& Schmitt 2004) and a deficit in the hot, hard X-ray component.  Non-accreting stars interestingly do not show the soft X-ray excess, and their absolute and fractional X-ray luminosities are a factor of $\sim 2-3$ higher than the CTTS (e.g., Preibisch et al. 2005, G\"udel \& Telleschi 2007). The origin of the X-ray emission is unclear, but data indicate that although mediated by accretion,  it arises largely from coronal magnetic activity. Accretion is believed to suppress X-ray activity by cooling coronal regions, thereby causing a reduction in the hot ($\sim 20-30$MK ) component emission and producing a soft excess  (e.g., G\"udel 2007,  Preibisch 2007). 

Due to the uncertainties in the origin of the soft X-ray emission and the lack of a simple method to adequately model the changes in the spectral character of the X-ray emission as the accretion rate declines, we neglect this soft component for most our models. 
Our default spectrum is more characteristic of a non-accreting weak-line T Tauri star (e.g., Feigelson \& Montmerle 1999), although  we do consider one  test case where we include a soft excess.   Our justifications are as follow.  Soft X-rays are easily absorbed  and will not penetrate the disk wind during early stages of disk evolution when accretion rates are high and when they are likely to be a significant component of the X-ray spectrum (ECD09). As we will show subsequently, 
FUV photoevaporation already depletes the disk mass substantially before the accretion rates and accompanying wind column densities are low enough for the soft X-rays (and EUV) photons to irradiate the disk surface.  As the disk dissipates and the accretion rate further declines, it may very well be that the X-ray spectrum is no longer mediated by accretion and that the emergent stellar X-ray flux is more representative of a wTTS, the template for our adopted  standard ``hard'' X-ray spectrum. 

We consider a soft X-ray excess for one disk model to probe the effects of the X-ray spectrum on disk photoevaporation. We note that some disks with low accretion rates, such as the face-on disk around TW Hya, show evidence of a soft X-ray excess (Kastner et al. 2002).  Additionally, we are interested in the possibility that soft X-ray photons may be significant in creating gaps in the disk. Soft X-rays can penetrate the disk wind (at late epochs) and heat dense gas to high temperatures ($\sim$ few 1000 K) which may  potentially result in higher photoevaporation rates, at least compared to rates produced by EUV  photons and \app 1keV X-rays (ECD09).  The ionization and heating rates are higher for soft X-rays due to the higher absorption cross-section for X-rays at low energies.  (The X-ray absorption cross-section $\sigma(E) \sim 10^{-22} (E/1{\rm keV})^{-2.6}$ cm$^{-2}$, see Wilms et al. 2000, GH08).  Higher ionization fractions also result in higher heating efficiencies but fewer secondary ionizations for X-rays (Shull \& van Steenberg 1985).  In GH08 and GH09, we did not achieve heating efficiencies higher than 10-20\% in the atomic gas, as  the ionization levels were not high enough for our chosen X-ray spectrum, with a peak at 2keV.  For our soft X-ray test model in this paper, we simulate the soft X-ray excess by assuming a spectrum that continues to rise at low energies.  Specifically, we assume that  $L_X(E) \sim E^{-1}$, for $0.1<E<2$keV. At higher photon energies, we retain the previous power law. This results in a overall spectrum somewhat similar to that observed for the soft-excess sources (Kastner et al. 2002, Stelzer \& Schmitt 2004).  We note that the adopted soft X-ray spectrum puts equal energy in logarithmically spaced intervals of $\Delta\nu$ or $\Delta E$. 

 \paragraph{EUV}
  The EUV luminosities of  stars are very poorly known and at best indirectly determined (e.g., Bouret \& Catala 1998, Alexander et al. 2005).  EUV photons are easily absorbed by low column densities of neutral gas ($\sim 10^{18} {\rm cm}^{-2}$),  and hence stellar EUV luminosities are very difficult to measure at typical star-forming region distances. 
  For low and intermediate-mass stars (M$_* <$ 3 \ms),
  we assume that the  EUV luminosity scales similarly with the chromospheric FUV and X-rays, and has a similar strength, $ L_{EUV} \sim 10^{-3} L_*$.  For M$_* >$ 3 \ms, the
  EUV luminosity predominantly arises from the photosphere.   As described earlier, we use data from Parravano et al. (2003) to estimate the EUV luminosities for more massive stars. 
   A knowledge of the EUV spectrum is not essential for calculating the EUV-induced photoevaporation rates. The temperature of EUV heated regions is nearly constant at $\sim 10^4$K, and the photoevaporative mass loss rates depend only on the total photon luminosity, which is $\sim 5\times 10^{40} $s$^{-1}$ for $L_{EUV}=10^{-3}$L$_{\odot}$. 

Other stellar parameters such as radius and temperature are taken from the pre-main-sequence tracks of Siess et al. (2000). We assume that the stellar optical spectrum is a blackbody at the assumed effective temperature. Table~\ref{table1} lists the stellar parameters for our different models. There is a minimum value for $L_X$ and $\Phi_{EUV}$
at $\sim 3$\ms\  due to a drop in chromospheric activity as convection ceases near the stellar surface (e.g., Feigelson et al. 2002). We do not list $L_{FUV}$ because of the time-dependent term due to accretion onto the star. However, for a 1\ms\ star, $L_{FUV}\sim 10^{-3}-10^{-2}$\ls.

\subsection{Disk Parameters and Model Assumptions}
Our disk model consists of gas and dust  that are well-mixed for simplicity. We assume that the initial disk mass ($M_{disk}$) scales with the mass of the central star as indicated by observational studies (e.g., Andrews \& Williams 2005, 2007) and specifically choose  $M_{disk} (t=0)  = 0.1 M_*$.  Our choice of initial disk mass  corresponds to that of a marginally gravitationally unstable disk (Pringle 1981). 
Although median disk masses inferred from dust emission are lower, \app \pow{3}{-2}\ms \ for the youngest objects (Andrews \& Williams 2005), they may have been systematically underestimated (e.g., Andrews \& Williams 2007, Hartmann et al. 2006).  It is also quite likely that disk masses just after collapse are higher, but gravitational instabilities may dominate and drive disk evolution at these early epochs and we ignore this phase of evolution.  The model disk extends from $0.1-200$ AU at $t=0$, and the surface density is initially a power-law with radius ($\Sigma \propto r^{-1})$.  We introduce a low surface density  ``tail" at larger radii (200-1000AU) for numerical purposes, which initially contains a negligible fraction of the disk mass. The disk quickly evolves under viscosity to a self-similar solution with $\Sigma(r)$ going approximately as $r^{-1}$ in the inner regions and with the outer radius spreading  with time (e.g. LBP74). 

Dust grains are assumed be of one size ($0.3\mu$m) for simplicity which corresponds to a factor of 10 reduction in opacity to optical or UV photons compared to  primordial interstellar cloud material.  The chemical composition is  astronomical silicates  and the absorption coefficients are calculated using data from the University of Jena database  (J{\"a}ger et al. 2003). Our dust model is as described at the beginning of \S 2 and  although simple,  it is adequate for the goals of this paper, which are mainly to show the time-dependent evolution of a viscously evolving disk that is subject to photoevaporation by X-rays and UV. We neglect settling and coagulation of dust in this initial treatment and will investigate the simultaneous evolution of the dust disk in a future paper.  

We assume that the viscosity parameter $\alpha$  is a constant throughout the disk. $\alpha$ here represents  a mean quantity averaged over $z$ at each $r$ and  specifies the local rate at which angular momentum is transferred (for a thin disk where the scale height $h(r) \ll r$).
%Numerical calculations of rotating magnetic disks indicate that a magnetorotational 
%instability can effectively yield a viscosity with $\alpha $ between 10-3 and 0.1 Brandenburg et al. 1995; Stone et al. 2000; Papaloizou & Nelson 2003). 
 Numerical simulations of the magneto-rotational instability, which is believed to be a plausible source of disk kinematic viscosity, yield an effective viscosity $\alpha$ between $\sim 10^{-3}-0.1$ for weakly ionized T Tauri disks (Brandenburg et al. 1995; Stone et al. 2000; Papaloizou \& Nelson 2003).  Hartmann et al. (1998) use observationally derived estimates of mass accretion rates and disk lifetimes in T Tauri disks to obtain $\alpha \sim 0.01$ .  We assume that $\alpha=0.01$ for a 1\ms \ star. 
 
 We further assume that $\alpha$ is proportional to the mass of the central star.  This scaling is motivated  by  observations of a correlation between the accretion rate and central star mass, $\dot{M}_{acc} \propto M_*^2$ (Muzerolle et al 2003, Natta et al 2004, Calvet et al. 2004,  Muzerolle et al 2005, Natta et al 2006). Although this correlation shows a very large scatter, it appears to be valid over a wide range of stellar masses.
 The physical origin of this correlation is unknown and many theories have been proposed in explanation, such as  mass-dependent stellar properties (e.g., Muzerolle et al. 2003; Mohanty et al. 2005; Natta et al. 2006b), initial core properties (Dullemond et al. 2006), Bondi-Hoyle accretion (Padoan et al. (2005) and scatter caused by spread in disk ages (Alexander \& Armitage 2006). It has also been argued that the correlation may be an observational bias and that the $\dot{M}_{acc} - M_*$ relation may in fact be less steep or even non-existent (Clarke \& Pringle 2006). Vorobyov \& Basu (2008) interpret the relation in the context of a non-viscous disk model, where gravitational torques drive accretion in the T Tauri disk resulting in $\macc \propto M_{disk}^{1.1}$, and further infer that $M_{disk} \propto  M_{*}^{1.3}$, resulting in a flatter correlation than is observed. In our $\alpha-$disk model, an $\alpha$ independent of stellar mass would give a linear dependence of $\macc$ with $M_*$, as $\macc \propto (\alpha M_{disk})$, and $M_{disk}\propto M_*$.  But a  linear scaling of $\alpha$ with the central star mass  recovers the observed correlation. We strive to reproduce the observational dependence mainly because our FUV luminosity is derived from the disk accretion rate. Our choice of the numerical value of $\alpha$ 
 ($0.01$ for a 1\ms\  star) has some observational and theoretical support (e.g., Hartmann et al. 1998, Papaloizou \& Nelson 2003), but not much is known about its scaling with stellar mass. The assumed linear dependence, i.e., $\alpha \propto M_*$, however, ensures that our accretion-generated FUV luminosities correspond closely with  measured values (e.g. IUE, Valenti et al. 2003, also see GH08).  
 
 Table 2 summarizes the basic parameters of our disk model. 
 
\subsection{Model Computation of Gas Temperature}
An accurate determination of the gas temperature as a function of spatial location in the disk is a very exacting task for the surface layers where gas and dust temperatures deviate ($T_{gas} \ne T_{dust}$).  Our earlier detailed, static thermo-chemical disk models were developed to solve this challenging problem, but their complexity makes them  very computationally intensive (GH08, GH09).  Attempting to integrate these gas disk models into a time-dependent code requiring full disk calculations at many ($\gg 10^4 $) timesteps is  not a feasible solution. We therefore develop simplified solutions to the gas temperature in the disk, whereby we use an interpolation scheme that  draws on results from our static thermo-chemical disk models. GH09 found that FUV photoevaporation acts mainly on the outer disk and depletes the disk of its mass reservoir. One of our main goals is to estimate disk lifetimes due to photoevaporation, and it is therefore important to determine the disk density and temperature structure accurately in these outer regions.  We find that simple approximations to the gas temperature, for example, assuming that the gas is entirely atomic, can fail in these outer regions. Molecular chemistry and cooling become important in the outer disk, and an accurate computation requires a detailed chemical network. We therefore avail of the data from  the full chemical network models(GH08)  for the gas temperature in the disk. 

 We identify six different parameters that influence the local gas temperature $T_{gas}(r,z)$ ---  the local gas number density $n (r,z)$, FUV flux $G_{FUV} (r,z)$,  X-ray flux $G_X (r,z)$, the dust temperature $T_{dust}(r,z)$, the dust opacity per H nucleus $\sigma_H$ and  the vertical column density to the disk surface $N_{up}(r,z)$. Gas heating and cooling rates depend on the gas density.  FUV and X-ray heating depend on the local (attenuated) fluxes  $G_{FUV} (r,z)$ and $G_X (r,z)$.  Gas is also heated  or cooled by collisions with dust  ($T_{dust}(r,z)$, $\sigma_H$  and $n$) and the  total gas cooling depends on the vertical column density through which the cooling line flux escapes ($N_{up}(r,z)$).   We then construct a look-up table using data from our many previous static model runs  that solve in detail for the thermal balance and chemistry (GH08, GH09). We calculated the disk structure for some additional cases so that the parameter space is approximately well represented for the range of anticipated values of these six quantities.

The gas temperature is determined using a $k-$nearest neighbour algorithm (Clarkson 1999, Beygelzimer et al. 2006) in our six dimensional {\it pseudo}-metric space and using an inverse distance weighting method (Shepard 1968) for interpolation between these $k$ neighbours ($k\sim10$). Because our six parameters are nearly independent and as our database is well populated, this method produces fairly accurate results. We tested our approximation scheme against several detailed computational models to verify their accuracy and errors  in the derived photoevaporation rates are estimated to be within $\sim 10$\%. 

\subsection{Model Computation of Disk Evolution}
\subsubsection{Evolution prior to gap formation}
We then solve for the disk surface density evolution as follows. We begin with the prescribed initial surface density distribution and let the disk viscously evolve. The mass accretion rate onto the central star is determined by the mass inflow from the first radial gridpoint and we assume that there exists an accompanying wind with a mass loss rate equal to $0.1\dot{M}_{acc}$.  We use the analysis of Hollenbach \& Gorti (2009) to calculate the column density in the wind given by
\begin{equation}
N_w = 0.1 \dot{M}_{acc} \left(\pi m_H v_w r_w (1+0.5f_w)\right)^{-1}
\end{equation}
where $v_w$ is the disk wind velocity and the wind originates from a cylindrical radius $r_w$ to $f_wr_w$. We assume typical values for these parameters, $v_w = 100$ kms$^{-1}$ and the wind base radius is a few stellar radii, $r_w=10^{12}$ cm and $f_w\sim1$
(e.g., Shang et al. 2007). 
The stellar X-ray and FUV flux is attenuated by this column density before it is incident at any spatial location $(r,z)$ on the disk, in addition to the attenuation through any intervening disk matter. We allow the disk wind to attenuate the FUV, even though the wind is probably located inside the dust sublimation radius (e.g., Shu et al. 1994) and may be considerably depleted of dust grains. The EUV photons are attenuated by the atomic H (and He) atoms at the base of the flow. The atomic H column $N_H$ can be considerably less than $N_w$. We perform a ``Stromgren-like" calculation (see Hollenbach \& Gorti 2009) to determine when the EUV photons ionize the column  $N_w$ and therefore penetrate to irradiate the disk.  Radiation may also be quenched by absorption in the accretion funnel (e.g.,  Alexander et al. 2005, Gregory et al. 2007), but we do not attempt to model this attenuation. The geometry of the accretion streams is quite complex, and moreover, probably  allows escape of at least some energetic photons that can irradiate the disk (e.g. Calvet \& Gullbring 1998).  In fact, young actively accreting stars are found to be quite FUV and X-ray luminous. As the disk evolves,  first FUV and then hard X-ray photons ($\gtrsim 1$keV)  penetrate the wind to shine on the disk.  At later stages when the accretion rate has declined substantially ($\macc$ \lapp $10^{-8}\mpy$), soft X-ray and EUV photons can penetrate the disk wind and  heat the disk surface. 

Once the FUV penetrates, the disk radiative transfer is invoked in order to calculate the vertical temperature and density structure of the gas and the photoevaporation mass fluxes from each radius.  We start with the dust temperature calculation followed by the gas temperature and vertically iterate each radial zone for hydrostatic equilibrium until convergence is reached. The resulting density and temperature profiles then yield the mass loss rate ($\dot{\Sigma}_{pe}$) at each radius, calculated using the analytical approximations of Adams et al. (2004) as described in GH09.  
Equation~{\ref{sdot}} is then advanced to the next time step to determine the surface density at the next time. The disk structure calculations are repeated to update $\dot{\Sigma}_{pe}$ at every timestep. 

\subsubsection{Direct photoevaporation after gap formation}
We treat gap formation and subsequent direct illumination of the disk using a procedure analogous to that of ACP06. We note again that once a gap forms, viscous accretion of the inner disk onto the central star proceeds rapidly leaving a central hole around the star. The star then shines upon the inner rim of a ``torus-like'' disk. 

\paragraph{EUV } ACP06 define EUV gap formation to have occured when the column density to the star at the midplane is low enough to allow penetration by ionizing photons.  We follow the same procedure and use a similar analysis for calculating the EUV photoevaporation rates after gap formation. 
A hole of size $r_h$ forms  in the disk when the accretion rate at
this radius falls below the photoevaporation rate, followed by rapid viscous draining of matter inside this radius. Stellar photons now directly irradiate the rim and photoevaporate the gas.  Following ACP06, the scale length in the radial direction is assumed to be $\sim f z_h$, where $z_h$ is the height of the neutral disk and $f$ is a parameter of order unity. We assume $f\sim 0.3$ to match the results from the hydrodynamical models of ACP06. The incident flux is balanced by the recombinations along the path length, and therefore at a position $(r_h, z)$ on the rim 
we have
\beq
\alpha_B n_e(z)^2 f z_h = F_{EUV} = { {\Phi_{EUV}}\over{4 \pi (r_h^2 + z^2)}}
\label{recomb}
\eeq
where $\alpha_B$ is the Case B recombination coefficient, $n_e$ is the electron density and $\Phi_{EUV}$ is the EUV photon luminosity. The mass loss rate off the rim is then given by  integrating the mass flux above and below the midplane,
\beq
\dot{M}_{pe,rim} = 2 \int_0^{z_h} \mu m_H n_e(z)  c_{II} 2 \pi r_h dz
\label{mpe}
\eeq
where $\mu m_H$ is the mean mass per particle  and $c_{II}$  the thermal speed of ionized gas. We can then derive the mass loss rate as
\beq
\dot{M}_{pe,rim}  \approx 4 \pi \mu m_H c_{II}  \left({ {z_h \Phi_{EUV}}\over{4 \pi f \alpha_B  }}\right)^{1/2} 
\eeq
where we have used the approximation $\sinh^{-1}(z_h/r_h) \approx z_h/r_h$. 
  ACP06 only consider absorption of EUV photons by the column of gas up to the ionization front (Eq.~\ref{recomb}). However, EUV photons are also required to ionize neutral hydrogen as it passes through the ionization front. This process can dominate the attenuation due to recombining H in the ionized gas at large disk radii. We choose $n_e$ to be the smaller of the values given by Eq.~\ref{recomb} and the relation ${{\Phi_{EUV}}/{4 \pi r^2}} =  n_e c_{II}$ (see Hollenbach et al. 1994).  
We  distribute this mass loss rate over a radial zone $r_i-r_o$ around $r_h$, to prevent numerical ``spikes". We set $\dot{\Sigma}_{rim} (r)$, 
assumed to follow a power law with radius, to be given by
\beq
\int_{r_i}^{r_o} \dot{\Sigma}_{rim} (r) 2 \pi r dr = \dot{M}_{pe,rim}
\eeq
Note that as we base our analysis on the total {\em mass loss} rate, the values of the parameters $r_i, r_o $ and the chosen power law for $\dot{\Sigma}_{rim} (r)$, 
 do not influence the total photoevaporation rates. These parameters are only  used to convert the mass loss rate from the inner rim to an equivalent  $\dot{\Sigma}_{pe}(r)$ for solving Eq.~\ref{sdot}. We check our results against the approximations provided by ACP06 using the results of detailed hydrodynamical models. Assuming the same viscosity law, $\Phi_{EUV}$ and by using their disk mass at the epoch of gap creation, we find that our above analysis of rim photoevaporation is in good agreement with their results.  

\begin{figure}
\plotone{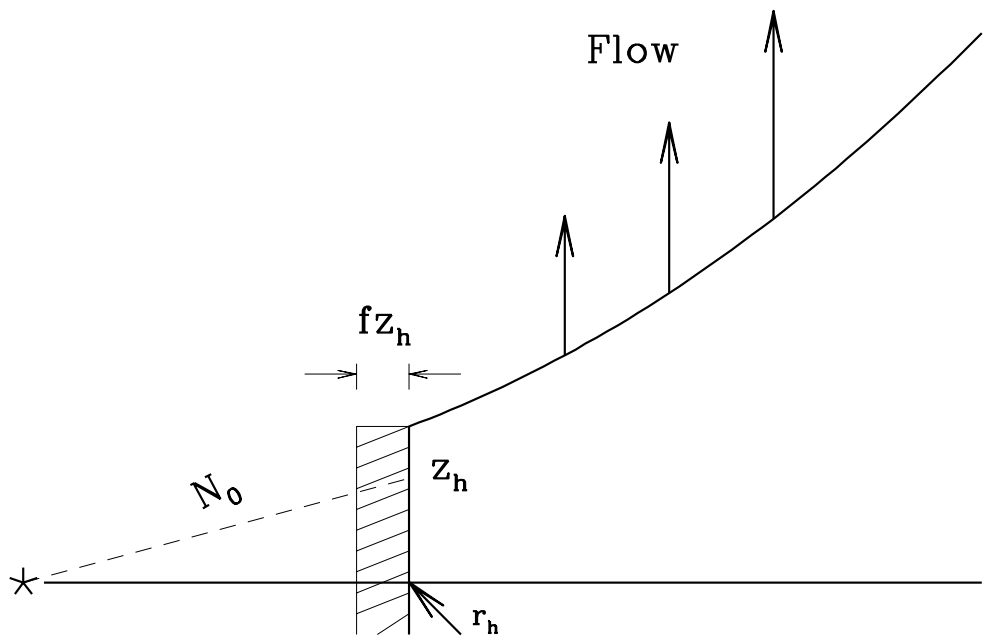}
\caption{ A schematic illustration of the rim structure and direct  FUV/X-ray photoevaporation following the  formation of a hole. The characteristic height $z_h$ is calculated as the height where the column density to the star is equal to $N_0$, where $N_0 \sim10^{22}$ cm$^{-2}$ corresponds to an optical depth of unity for X-rays and  FUV for our standard dust opacity. A radial column of $N_0$ through the scale length $f z_h$ is assumed to be set up in the subsequent flow for $z<z_h$, where $f\sim 0.3$.  }
\label{fig1}
\end{figure}

\paragraph{FUV and X-rays}
After the formation of the hole, FUV and X-rays photons are also incident directly on the inner wall of the disk.  We follow a similar procedure as above to calculate the mass loss due to rim irradiation by FUV/X-rays (Figure~\ref{fig1}).  At $r_h$, the vertical structure is calculated by the thermal balance calculations for every height $z$,  providing the density $n(r_h,z)$. We define the characteristic height $z_h$ here as the height at which the column density  $N_0$ to the star corresponds to an optical depth of unity for X-rays and FUV photons. 
We assume a scale length $fz_h$ in the radial direction (and again choose $f\sim0.3$ for consistency), and therefore photons are attenuated between $r_h-fz_h$ and $r_h$.  The attenuation column to the rim at each $z$ along $r_h$  is the minimum of $n(r_h,z) f z_h$ and  $N_0$. For our standard dust opacity, $N_0=10^{22}$cm$^{-2}$ for FUV photons. Our adopted gas absorption cross-section for X-rays (GH04) gives $N_0\sim10^{22}$cm$^{-2}$ for hard ($\gtrsim $1keV)X-rays and  $N_0=10^{20}$cm$^{-2}$ for soft ($\lesssim 0.2$keV) X-rays. We consider a range of attenuation columns ($N_0=10^{19}-10^{23}$cm$^{-2}$)  in our gas temperature calculations and calculate the photoevaporation rate $\dot{\Sigma}(r_h,z)$ as earlier described for surface FUV/X-ray photoevaporation. At each $z$ we use the value of $N_0$ that gives the maximum  $\dot{\Sigma}(r_h,z)$, although  $\dot{\Sigma}(r_h,z)$ is not particularly sensitive to $N_0$. Typically, the maximum occurs at $N_0\sim
10^{21}$cm$^{-2}$. The mass loss rate from the rim directed radially inward is then
\beq
\dot{M}_{r_h, FUV} = 2 \int_0^{z_h} \dot{\Sigma}(r_h,z) 2 \pi r_h dz
\eeq
which is again distributed over a radial zone bounded at $r_i$ and $r_o$. We assume for
simplicity that $\dot{\Sigma}_{rim}$ is a constant over this region. 

The total photoevaporation rate from the disk is then given by the sum of the direct photoevaporation rate from the rim as calculated above and the surface photoevaporation arising from the disk beyond the rim. We use the resulting net $\dot{\Sigma}_{pe}$ in Eq.~\ref{sdot} and continue solving for the surface density until the entire disk is dispersed. 

 \section{Results and Discussion}
 \subsection{Standard case (1\ms star)}
We first describe the results of our photoevaporation model for a $0.1{\rm M}_{\odot}$ disk around a $1{\rm M}_{\odot}$ star.   In order to isolate the relative importance of EUV, FUV, X-rays and viscosity in disk evolution, we consider several combinations of these four parameters and study the surface density distribution and disk mass as a function of time.  The different model disks are (a) only viscosity included,  (b) viscosity and EUV photoevaporation (c)  viscosity and FUV photoevaporation and lastly, (d) a disk with viscosity and EUV, FUV and X-ray photoevaporation.  We do not consider an isolated X-ray model, that is a disk with X-rays but no EUV or FUV.  This because the removal of FUV also removes molecular photodissociation. Consequently, there is strong molecular cooling which keeps the disk flat, cold and tenuous at the surface regions where X-ray photoevaporation can be important,  resulting in  very little mass loss. We discuss X-ray photoevaporation in further detail in section \S3.2 where we consider the effect of the X-ray spectrum on photoevaporation rates. 

\begin{figure}
\plotone{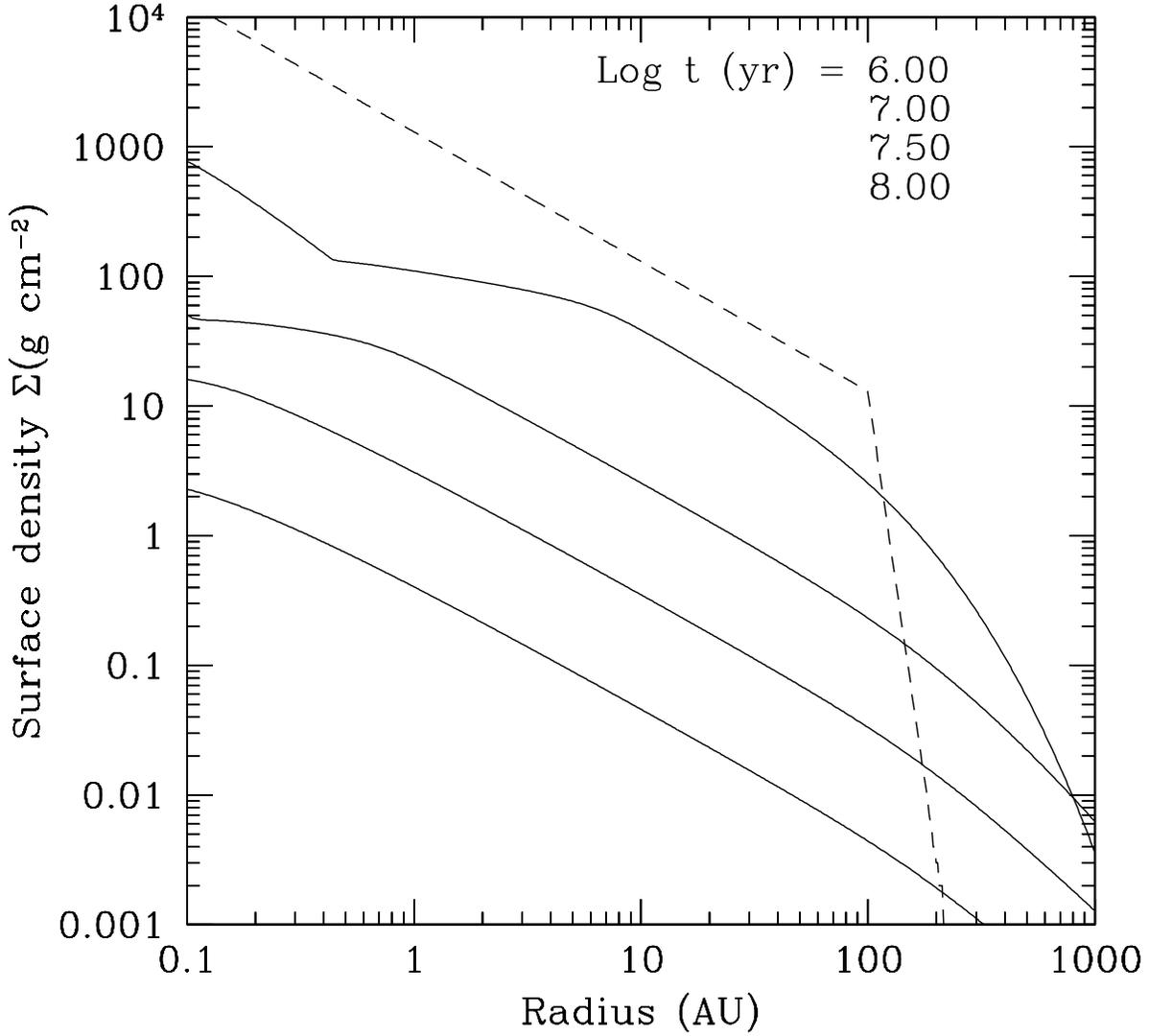}
\caption{Surface density($\Sigma$) distribution with time for a purely viscous disk ($\alpha=0.01$, initial disk mass $0.1M_{\odot}$) around a 1M$_{\odot}$ star. The dashed line shows $\Sigma$ at the start of the simulation, $t=0$.  $\Sigma$ with radius is shown for different instances of time indicated in the upper right hand corner. The surface density gradually decreases with time as the disk spreads, and the disk mass is $>10^{-3}$M$_{\odot}$at $10^8$ years. The disk is optically thick to stellar photons until
$\Sigma \lesssim 10^{-2}$ g cm$^{-2}$. }
\label{fig2}
\end{figure}

\paragraph{Viscous disk with no photoevaporation}
 In the first case (Fig.~\ref{fig2}), the disk gradually accretes onto the star and the surface density declines with time while it simultaneously spreads out in radius due to viscous evolution.  At $10^8$ years, almost 98\% of the initial disk mass has accreted onto the star, leaving a large, extended disk of $\gtrsim 10^{-3}$M$_{\odot}$. 
 Even though very little mass is left in the disk, the surface density at these relatively late epochs is still  well above current observational detection limits for both gas (CO) and dust emission and an ``optically thick" disk remains.  For a disk to be optically thick in gas or in dust, the vertical column density through the disk has to be greater than $\sim 10^{22}$ cm$^{-2}$, or $\Sigma \gtrsim 10^{-2}$ g cm$^{-2}$. Figure~\ref{fig2} shows that the surface density is higher  than this value even at $10^8$ years and the long disk lifetime is contrary to observations.  Therefore, a purely viscously evolving disk clearly does not explain the observed rapid dispersal and short lifetimes of protoplanetary disks.

 One can see that the disk spreads outward as it accretes onto the star  
and loses mass. It thereby roughly follows the behavior of the  
analytic solutions of LBP74, except that in the inner  
regions one can notice some kinks. First of all there is a region  
where the surface density becomes almost flat. This is because in this  
region the viscous heat production near the midplane is dominating  
over the irradiation. The temperature increases much more rapidly with  
decreasing radius than in the irradiation-dominated outer disk region.  
Since from standard accretion disk theory one can derive that for  
steady or semi-steady disks, at any given instant in time,  $\Sigma(r) \cdot T(r)\cdot   
r^{3/2}$ is constant with radius, a steeper temperature slope means a shallower  
surface density slope. Hence the near-constant surface density region.  
Then further inward one sees that the surface density suddenly follows  
a very steep slope: $\Sigma\propto r^{-3/2}$, meaning that in this  
region the midplane
temperature is constant with radius. This is because in this region  
the dust starts to evaporate. But since the dust is also responsible  
for most of the opacity, the evaporation of dust also lowers the  
midplane temperature. What happens in this region is that the dust  
acts as a kind of "thermostat", fixing the midplane temperature at  
precisely the dust evaporation temperature. If more dust was to  
evaporate, the temperature would drop below the dust evaporation  
temperature, allowing dust to recondense. Oppositely, if too much dust  
is present, the midplane temperature rises, and more dust is  
evaporated. This thermostat effect causes the surface density slope to  
be steep ($\Sigma(r) \propto r^{-3/2}$) in that very inner region.
 
\begin{figure}
\plotone{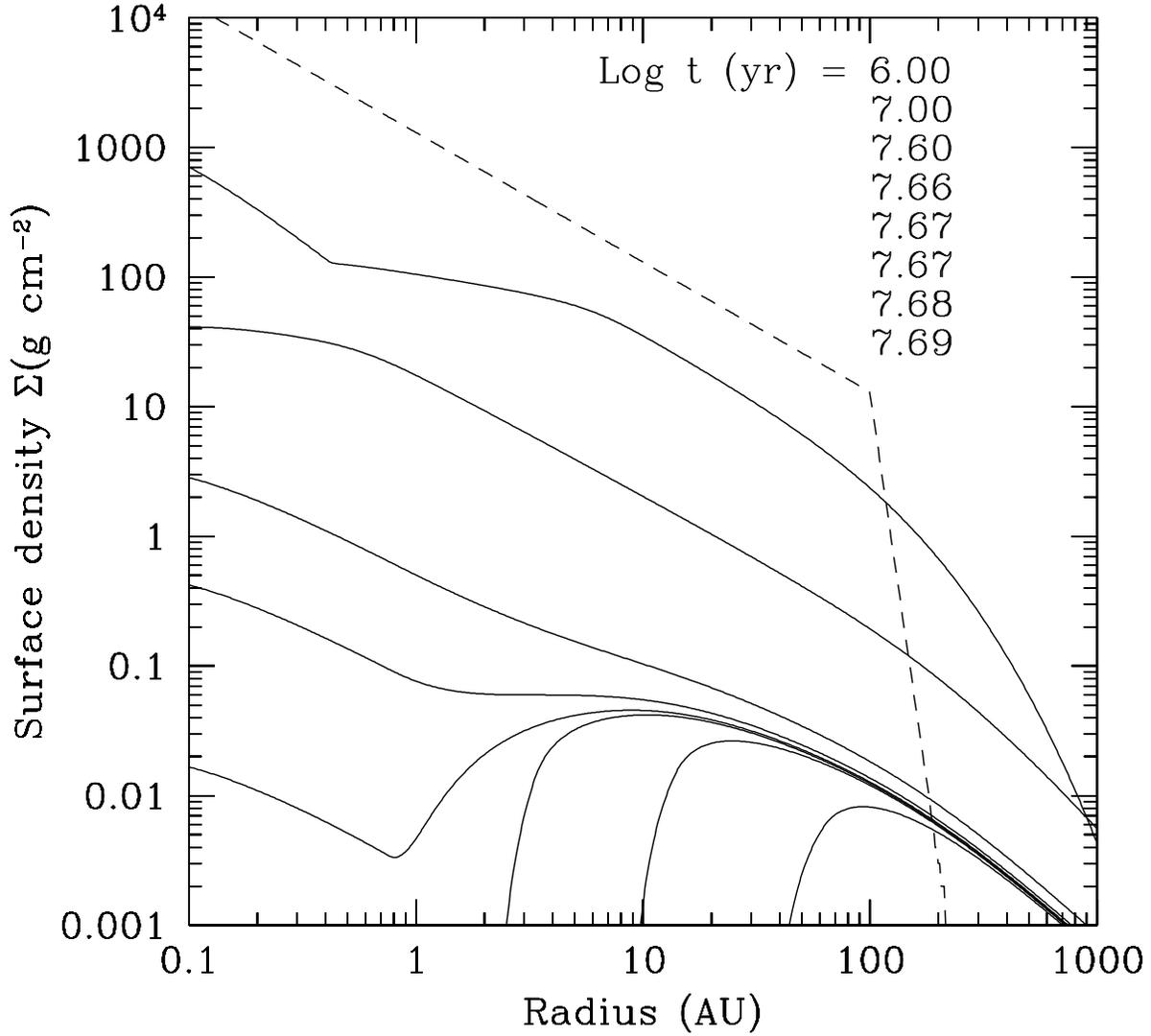}
\caption{The evolution of the surface density distribution for the same  disk/star system as Fig.~\ref{fig2},  but which is irradiated only by EUV photons ($\Phi_{EUV}=5\times 10^{40}$s$^{-1}$). The disk loses mass gradually  and when the accretion rate is low enough, EUV photons burn a gap in the inner disk at $\sim 4.7 \times 10^7$ years. The disk is then eroded by direct illumination of the inner gap by EUV photons.}
\label{fig3}
\end{figure}

\paragraph{Viscosity and EUV} Figure~\ref{fig3} illustrates the evolution of a disk subject to EUV photoevaporation and viscosity, as was first proposed by Clarke et al. (2001). At initial stages, disk accretion rates are high ($\gtrsim 10^{-8} {\rm\ M}_{\odot} {\rm yr}^{-1}$) and the column density in the disk wind is too high for penetration by EUV photons. The disk evolves purely viscously during these epochs and the evolution of the surface density distribution is similar to that in Fig.~\ref{fig2}.  After the onset of EUV irradiation, the  disk surface is heated to $10^4$K and photoevaporates with  mass loss rates that are initially much lower than the disk accretion rate.  Viscous accretion replenishes disk mass at a given radius faster than  EUV photoevaporation can deplete mass and therefore for almost $\sim 4\times 10^7$ years, the disk evolves almost as in the purely viscous case, albeit with a small additional mass loss due to  the photoevaporative wind.  When EUV photons can remove disk mass at a given radius faster than it can be replenished by viscosity, a gap opens at that radius,  as was first noted by Clarke et al. (2001). 
  The rate of decline of surface density $\dot{\Sigma}_{pe}$ peaks at a critical radius $r_{cr} \sim 0.1-0.2 r_g$ (Lifman 2003, Adams et al. 2004, Font et al. 2004) where $r_g$ is the gravitational
 radius in the disk defined by $r_g = GM_*/c_s^2$($\sim 7 $AU for a 1M$_{\odot}$ star, see Hollenbach et al. 1994).  The mass loss rate due to photoevaporation peaks at $r_g$ 
 ($\dot{M}_{pe} \propto r^{1/2} $ for $r<r_g$) and  it is here that the EUV photoevaporation rate is expected to first exceed the steadily declining mass accretion rate $\dot{M}_{acc}$  through the disk.  If $\dot{M}_{acc}$ were approximately constant with $r$ as is true for a steady-state viscous accretion disk, this would result in gap formation at $r_g$.  However, for a photoevaporating disk, the removal of mass at $r \sim r_g$
results in  $\dot{M}_{acc}$ ($\propto \Sigma(r)$) decreasing as the radius decreases for $r \lesssim r_g$.  The net result is that the gap first forms closer to $r_{cr}$, where 
$\dot{\Sigma}_{pe}$ peaks.  

Disk evolution is relatively rapid once the gap opens, at $\sim 4.6\times 10^7$ years for the adopted parameters. Viscous timescales in the inner disk being very short $\sim 10^5$ years, the inner disk depletes rapidly onto the central star, forming a ``hole''.  From this epoch onwards, EUV directly illuminates the  exposed rim of the outer disk, increasing the photoevaporation mass loss rate, to deplete the outer disk on short timescales (also ACP06). The survival time for our model disk after gap opening is  $\gtrsim 10^6$ years, contrary to the much shorter, $\sim 10^5$ year timescale found by ACP06. The main differences are due to our lower $\Phi_{EUV}$ and  due to their choice of a much lower initial disk mass. Observations of [NeII]12.8$\mu$m emission from disks indicates that $\Phi_{EUV}$ is unlikely to be as high as that adopted by ACP06 ($10^{42}$ s$^{-1}$ ) and favour our lower estimate (Hollenbach \& Gorti 2009, Pascucci \& Sterzik 2009). The initially small disk mass of the ACP06 models results in a lower remaining mass in the outer disk after gap formation (the disk has not expanded as much since the gap forms more quickly) and thereby also shortens the derived disk lifetime. We find that the ``EUV switch" model is not quite as abrupt, and that the disk may last longer after the creation of the inner hole, $\gtrsim 10^6$ years for expected EUV fields.  
Moreover, the gap opening timescale of a disk subject only to EUV photoevaporation is large, $\sim 5\times 10^7$ years.  In this model disk, $\sim$ 92\% of the disk mass is accreted onto the star and $\sim$ 8\% is lost due to EUV photoevaporation.   We conclude that EUV photoevaporation may not be important  for disk dispersal. 

\begin{figure}
\plotone{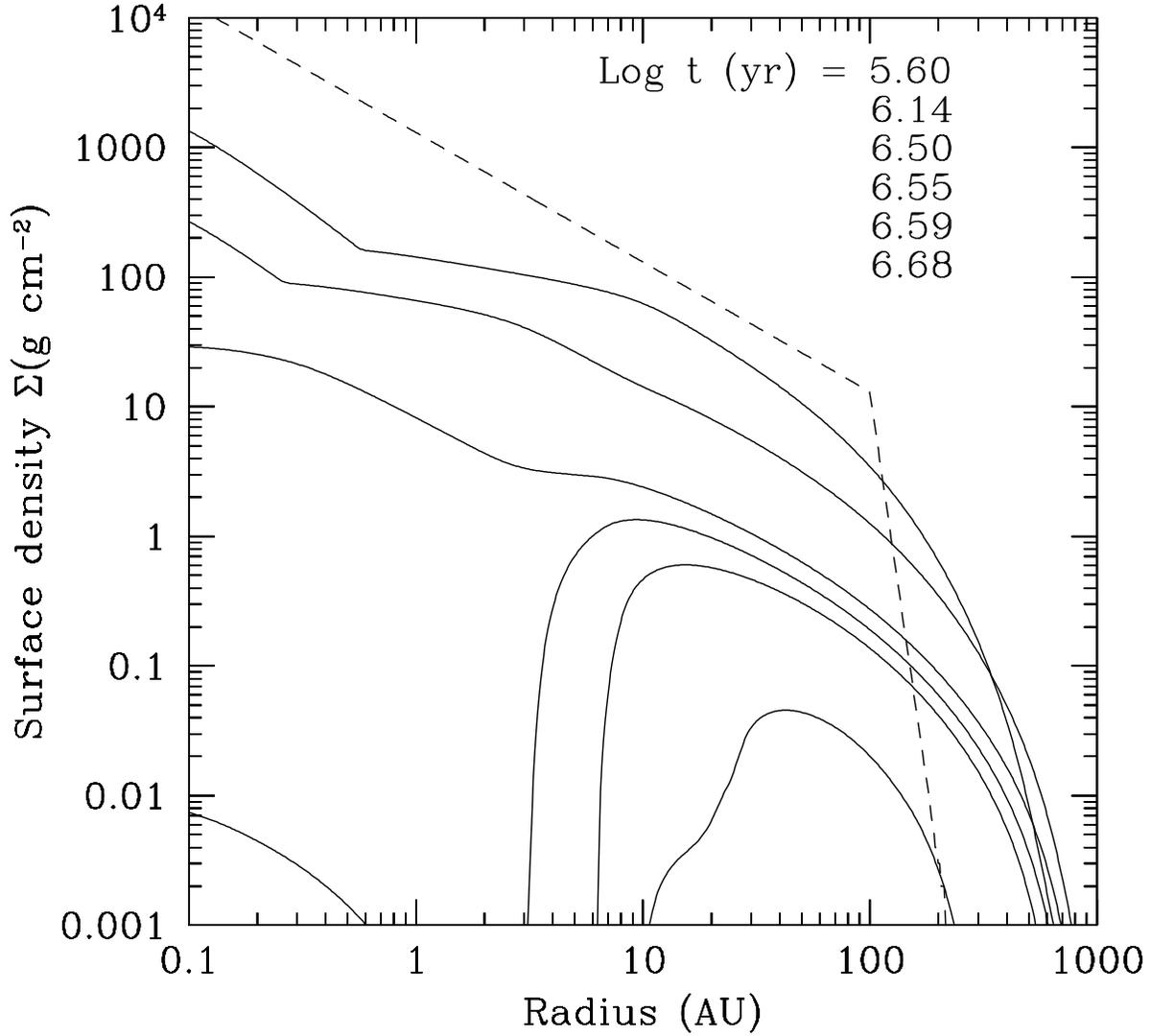}
\caption{ The evolution of the surface density distribution for  the same  disk/star system as Fig.~\ref{fig2} but which  is irradiated only by FUV photons ($L_{FUV}=10^{-2}-10^{-3} L_{\odot},\Phi_{EUV}=0, L_X=0$). The disk loses mass rapidly due to a combination of accretion and FUV photoevaporation and at $\sim 3.5\times 10^6$ years, FUV photons burn a gap in the inner disk. The disk is then photoevaporated by direct illumination of the inner gap as the outer disk continues to deplete. The  remaining torus-like disk is eroded at both the inner and outer regions, while the intermediate regions survive the longest. }
\label{fig4}
\end{figure}

\paragraph{Viscosity with FUV}
Figure~\ref{fig4} shows the evolution of a disk with only FUV  photoevaporation, i.e., we assume that $\Phi_{EUV}=0$ and that $L_X=0$ for this model. 
A considerable fraction of the accretion luminosity is radiated at FUV wavelengths, and at early epochs, the accretion rate  and therefore FUV flux from the star is high. The column density in the disk wind is also high, but FUV photons can penetrate through pure gas columns of $\sim 10^{24}$cm$^{-2}$. We do not deplete the dust in the wind, however, and assume that the dust opacity is the same as that in the disk. Therefore,  FUV is relatively unattenuated by the wind only when the wind column densities $N_w \lesssim 10^{22}$cm$^{-2}$ or when the disk accretion rate is $\lesssim 10^{-7} \mpy$. 

FUV photoevaporation shortens the disk lifetime considerably, as can be seen from Fig.~\ref{fig4}. The disk surface density decreases with time rapidly at all radii, but photoevaporative mass loss rates are higher at larger disk radii. This causes a steepening of the disk surface density profile in the outer disk from $r^{-1}$ for the earlier cases where viscosity prescribes evolution, to $r^{-2}$ or steeper. Due to the steep surface density slope caused by FUV photoevaporation, the matter in that region is in fact moving outward due to viscous torques, instead of inward. Therefore more matter from the inner/intermediate disk regions is ``fed"  into regions where FUV mass loss is high,  which accelerates the FUV photoevaporative destruction of the disk considerably. In other words, the surface density in the inner disk drops not only because of viscous accretion onto the star, but also due to spreading to large radii where it photoevaporates. 

An important result is the formation of a gap by FUV photons in the inner disk, even in the absence of EUV and X-rays. We had speculated on the likelihood of such a gap in our earlier static analysis (GH09) and we find that FUV photons are indeed capable of forming gaps in disks, even as the accretion rate and accretion luminosity decline in the disk.  In order to show how the relative values of the local accretion rate and local mass loss rate result in the formation of a gap at a given radius and lead subsequently to an inner hole, we  introduce a quantity that we call  the radial photoevaporation rate as $\langle \dot{M}_{pe}\rangle = 2 \pi r^2 \dot{\Sigma}_{pe}$. (We note that the actual mass loss rate from a radial annulus around $r$ is given by $2\pi r \dot{\Sigma}_{pe} dr$, which is smaller than the quantity  $\langle \dot{M}_{pe}\rangle$.) We also calculate the radial mass accretion rate  $\dot{M}_{acc}$ from the surface density and midplane temperature at a given epoch and compare this with $\langle \dot{M}_{pe}\rangle$ as a function of radius. 

\begin{figure}
\plotone{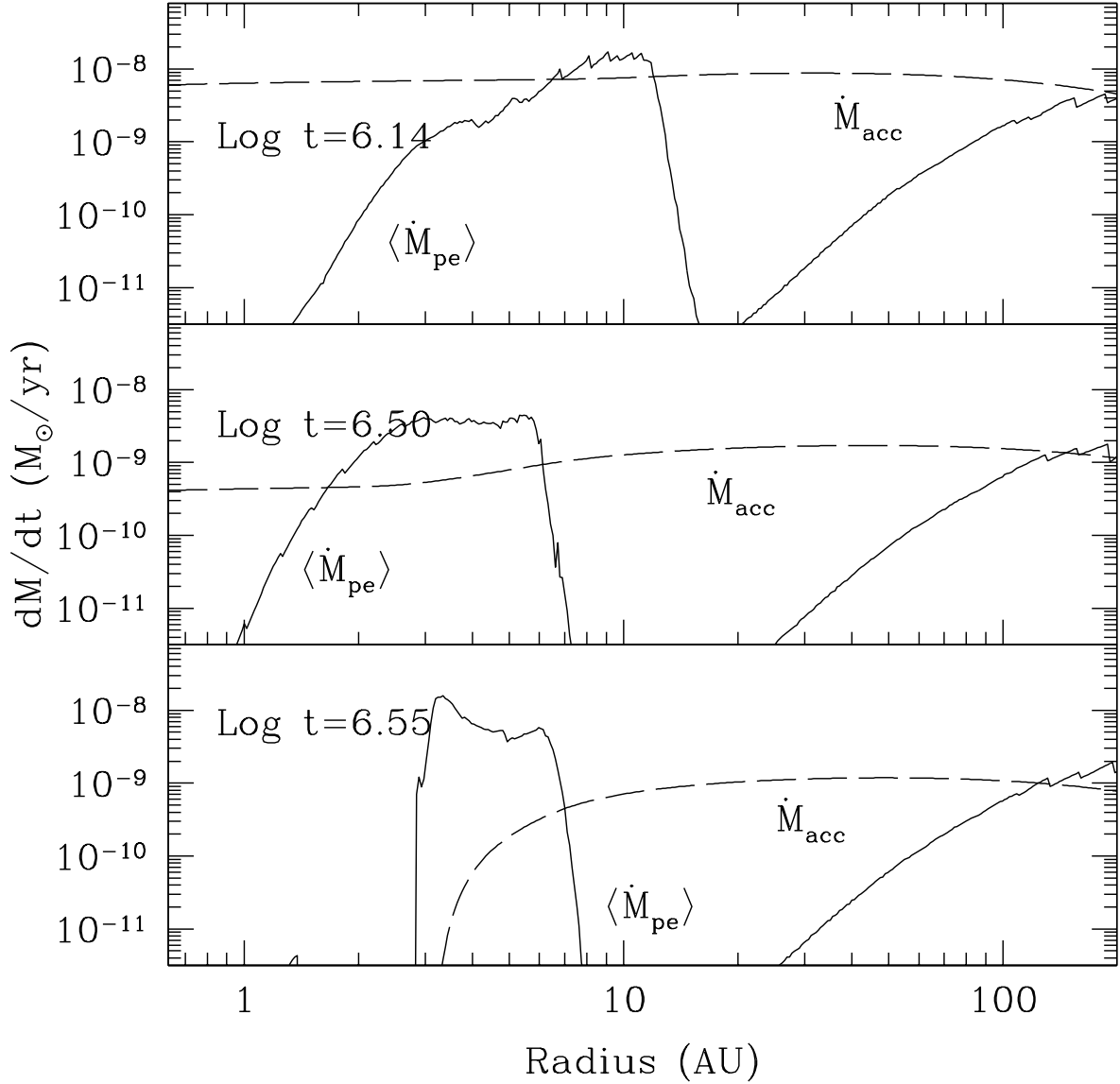}
\caption{ The radial photoevaporation rate ($\langle \dot{M}_{pe} \rangle=2\pi r^2\dot{\Sigma}_{pe}$, solid line) and mass accretion rate ($\dot{M}_{acc}$, dashed line) as a function of radius at three different epochs for the disk irradiated only by FUV photons. The gap forms when $\langle \dot{M}_{pe} \rangle  >\dot{M}_{acc}$ at $\sim 3\times 10^6$ years (second panel) at $r\sim 2-6$ AU. After the gap opens, direct illumination of the inner rim increases $\langle \dot{M}_{pe} \rangle $ by a factor of $\sim 3-5$. Note the steep decline in  the photoevaporation rate in the intermediate regions of the disk ($r\sim 10-30$AU) where disk mass survives the longest. In the outer regions of the disk, $\dot{M}_{acc} \gtrsim \langle \dot{M}_{pe} \rangle $ and viscosity continually feeds mass into these photoevaporationg zones, thereby aiding disk dispersal.
}
\label{fig5}
\end{figure}

 Fig.~\ref{fig5} shows the mass accretion rate(dashed line) and the radial photoevaporation  rate (solid line) in the disk before, during and after gap formation. The mass loss rate in the first panel of Fig.~\ref{fig5} is seen to first increase with disk radius, drop sharply at intermediate radii and rise again towards large disk radii. This behaviour is due to the higher gas temperatures ($\sim$ few 1000 K) in the inner disk due to efficient FUV and X-ray heating. At intermediate regions, the gas temperature begins to fall and the gravitational field is still fairly strong, i.e., $r_g \gg r$ in 
Eq.~\ref{dsonic}, resulting in low mass loss rates here. We note that the flows induced by FUV/X-rays at these regions of the disk are predominantly subsonic (Adams et al. 2004).  At larger disk radii,  gravity is weaker, but gas temperatures do not fall quite as rapidly, resulting in relatively high $\dot{\Sigma}_{pe}$ combined with increased disk surface area to produce enhanced mass loss from these regions (also see GH09). The second panel shows how the  gap first opens when $\dot{M}_{acc}$ first becomes lower than $\langle \dot{M}_{pe}\rangle $,  at $\sim 2$ AU and at $\sim 3\times 10^6$ years. There is direct illumination of the rim after the hole forms, and $\langle \dot{M}_{pe} \rangle $ increases here after the inner hole is created.   The entire disk is dispersed in $5\times 10^6$ years, when the surface density at all disk radii is zero.
The disk again lasts for $\gtrsim 10^6$ years after the hole is formed.  Note that although FUV photoevaporation rates are much higher than EUV-induced mass loss rates (by factors of $10-100$), the survival time for the disk after the hole forms is nearly the same as the pure EUV case. This is because of the larger disk mass($\sim 4\times 10^{-2}$\ms) at the gap creation epoch for the FUV case, where the gap is created at earlier stages of accretion.  Additionally, the gas temperature attained by FUV heating declines with distance to the star and the direct illumination of the rim only enhances the mass loss rates significantly in the inner disk regions. The outer disk continues to lose mass from the surface rapidly, and the disk evolves into a ring-like structure. The intermediate regions of the disk, where gas temperatures attained are not very high and where the stellar gravitational field is moderate,   survive for the longest periods. The total amount of disk mass accreted onto the star for this model disk is $\sim$ 53\%, while $\sim$ 46\% is removed by FUV photoevaporation. 
 
\begin{figure}
\plotone{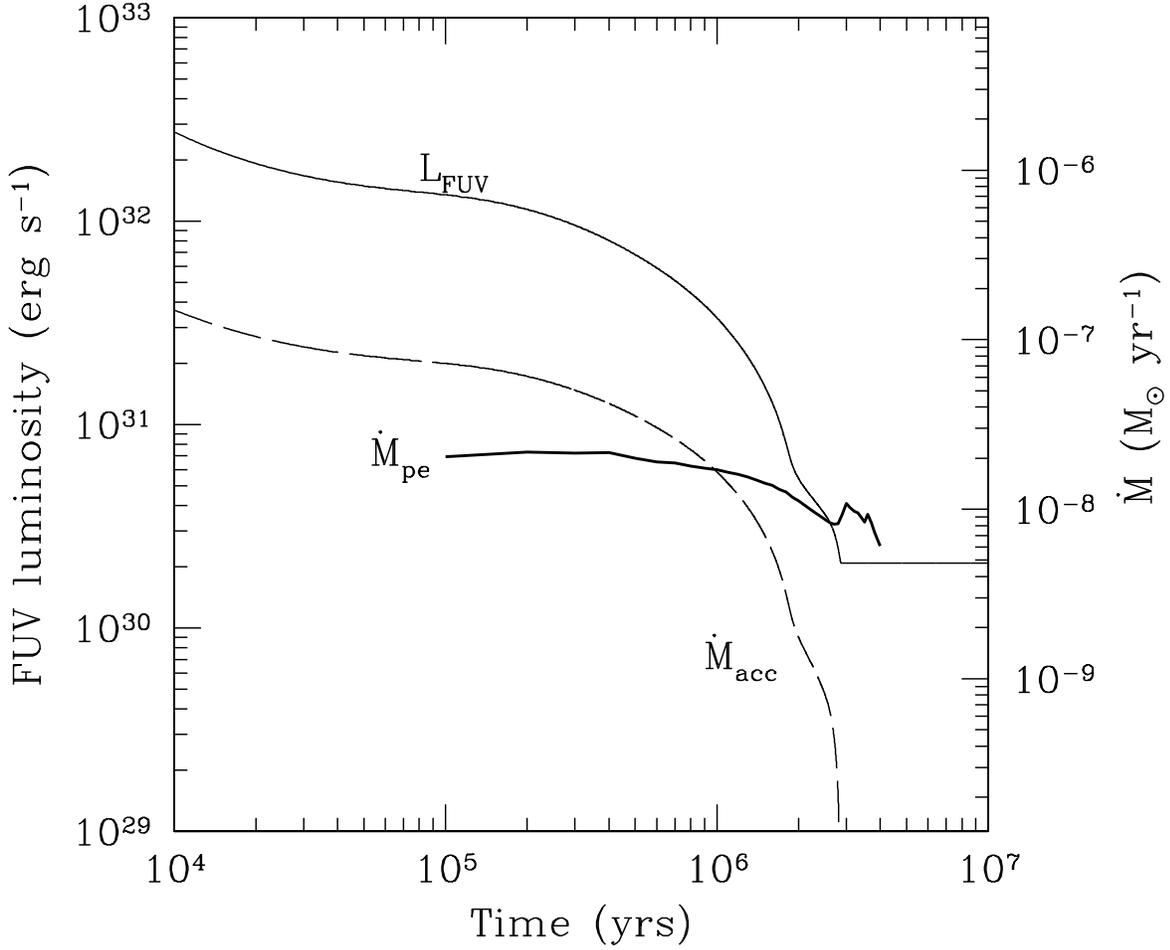}
\caption{The FUV luminosity as a function of time (solid line) as derived from the
instantaneous mass accretion rate (dashed line) for our standard model (\mst$=$1\ms) that
assumes $\alpha=0.01$ and includes EUV, FUV and X-ray irradiation. The FUV luminosity saturates at the chromospheric value when the mass accretion rate declines sharply after the formation of a gap in the disk. The total photoevaporative mass loss rate in the disk
$\dot{M}_{pe}$ is also shown as a function of time (thick solid line). $\dot{M}_{pe}$ is high throughout disk evolution, $\gtrsim 10^{-8} \ \mpy $.  }
\label{fig6}
\end{figure}

\begin{figure}
\plotone{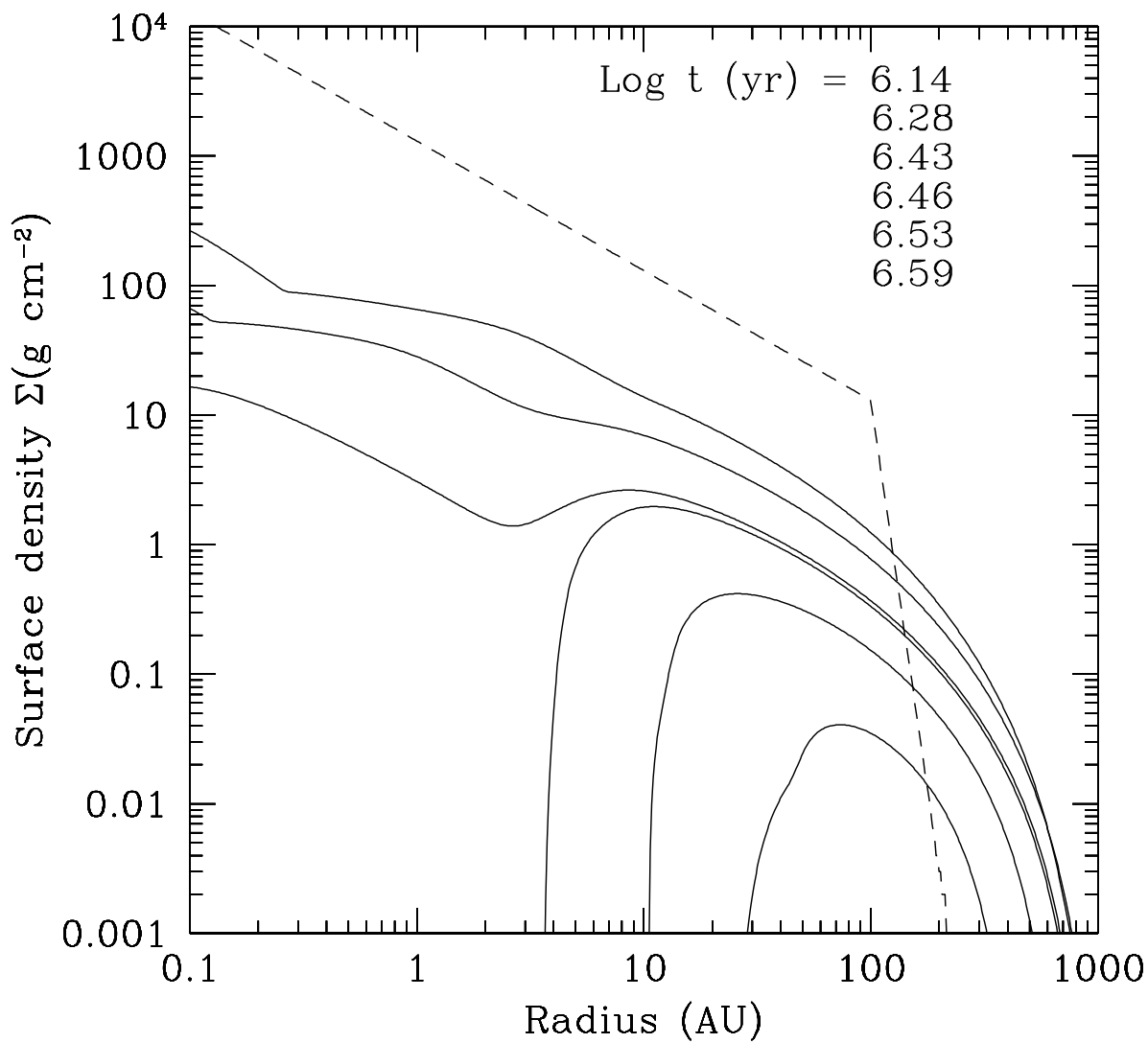}
\caption{The evolution of a viscous disk with EUV, FUV and X-ray irradiation. The disk is seen to be short-lived, and is destroyed in $\sim 4 \times 10^6$ years. Photoevaporative mass loss is enhanced significantly due to FUV and X-rays.  There is a gap formed at $\sim 2-4$ AU by FUV/X-rays, and the disk survives for $\sim 10^6$ years after gap creation.}
\label{fig7}
\end{figure}

\paragraph{Viscosity with EUV, FUV and X-rays} 
We next consider the evolution of a viscous disk that is subject to EUV, FUV and X-ray photoevaporation.  The disk  initially evolves viscously without photoevaporation, as there is no wind penetration of EUV, FUV or X-rays. As the accretion rate declines, at $\sim 10^5$ years, FUV and hard 1keV photons begin to irradiate the disk.
Figure~\ref{fig6} shows the FUV luminosity (solid line) as derived from the instantaneous mass accretion rate onto the star (dashed line)  as a function of time, and $L_{FUV}^{acc}$ is found to be of the order of $\sim 10^{-2} - 10^{-3}$ L$_{\odot}$ over the lifetime of the disk.  Accretion contributes almost entirely to the total FUV flux at early times,  and at late stages when $\dot{M}_{acc}$  drops steeply, chromospheric FUV dominates.  Also shown in the figure is the photoevaporation rate ($\dot{M}_{pe}$) in the disk ($\dot{\Sigma}_{pe}$ integrated over radius) as a function of time.  Photoevaporation begins with disk illumination, at $\sim 10^5$ years, and the mass loss 
rates are high $\sim$ few $10^{-8} \mpy$, through the entire evolutionary period. There is a slight decline in $\dot{M}_{pe}$ as the FUV luminosity declines, and a slight increase at $2.7\times 10^6$ years when a gap opens in the disk. 
  The evolution of this disk is slightly faster that the pure FUV case due to the addition of X-ray heating. Figure~\ref{fig7} shows the surface density evolution in the disk and the formation of a gap at $\sim$ 3 AU. The entire disk is dispersed in $4\times 10^6 $ years.  We find that the evolution of the disk is driven by the depletion of the mass reservoir in the outer disk where FUV photoevaporation rates are highest (also see GH09). The addition of X-rays only moderately enhances photoevaporation rates, a result we had earlier found in our static analysis (GH09).   However, a softer X-ray spectrum induces higher photoevaporation rates than obtained for this model, as we will show later.

\begin{figure}
\plotone{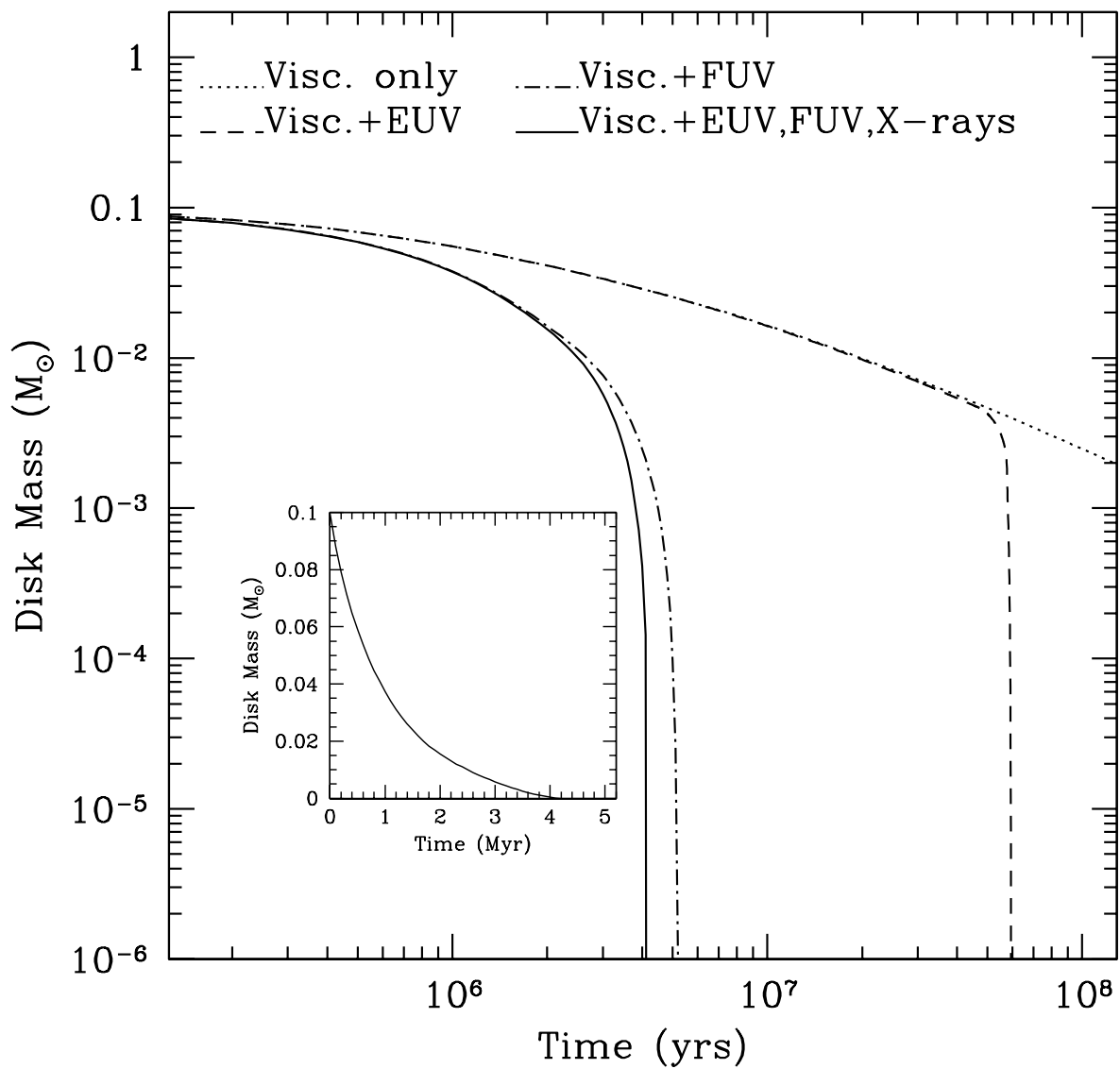} 
\caption{Disk mass as a function of time for all the models, (a) viscosity (green, dotted line), (b) viscosity and EUV (red, dashed), (c) viscosity and FUV (blue, dot-dashed) and (d) viscosity with EUV, FUV and X-rays (black, solid). The inset is a linear plot of the disk mass with time for this case.}
\label{fig8}
\end{figure}

Figure~\ref{fig8} shows the mass of the disk as a function of time for the four cases discussed above.  It is seen that the disk mass drops very sharply at late stages of evolution in the disks with photoevaporation, and we define this time as the disk lifetime. In contrast to the definition in our earlier work (GH09) where $\tau$ was a characteristic depletion timescale (the time for disk mass to drop by $e$), here we define $\tau_{disk}$ as the time for the disk to {\em completely} disperse. We find that viscously accreting disks around 1\ms\ stars and subject to FUV, EUV and X-ray photoevaporation can survive for $\sim 4 \times 10^6$ years before they are destroyed. In contrast, disks lifetimes of purely viscous disks with no photoevaporation are $\gg  10^8$ years (they never completely disperse, but become optically thin), and with only EUV photoevaporation are $6\times 10^7$ years. 

 To summarize our results, we find that FUV luminosities generated by accretion are capable of destroying disks rapidly. Approximately half of the initial $0.1M_*$ disk mass is viscously accreted by the star, while the remaining half is lost due to photoevaporation, mainly by accretion-generated FUV photons.  In the present analysis, we find that FUV  can also  create gaps in the disk before EUV photons can affect the evolution of the inner disk.  FUV/X-rays and EUV, however, open gaps in different regions of the disk (EUV at $\lesssim1$AU; FUV/X-rays at $2-5$AU)  due to the different temperatures to which they can heat the gas, which may hold important implications for any ongoing planet formation in the disk. The FUV/X-ray mass loss rates are higher than  EUV mass loss rates; therefore, the FUV/X-ray gaps open sooner and there is more disk mass in the outer disk when they do form. We calculate $\dot{M}_{pe} \sim 10^{-10}\ \mpy$ for EUV photoevaporation, and $\sim 10^{-8}\  \mpy$ for FUV photoevaporation.   Our main finding is that FUV-induced photoevaporation rates are a factor of $\sim 100$ higher than EUV-induced rates.  Disk lifetimes are therefore not affected by the EUV field,  with FUV (aided by X-rays) driving the mass evolution of the disk. We remind the reader that the photoevaporation process only removes the gas and smaller dust particles (sub-mm sized), leaving any larger objects (cm-sized or greater) behind in the disk. These larger solids may eventually form planetesimals and planets in the disk.  

Recent  ground-based, high-resolution observations of the [NeII]12.8$\mu$m line in a sample of disks  (Pascucci \& Sterzik 2009) support our  scenario of disk photoevaporation. These authors find that the line profiles of the [NeII] emission are consistent with ongoing EUV photoevaporation (Alexander 2008) for the more evolved transition disks (with low $\macc$) in their sample. Furthermore,   derived values of the EUV luminosity are low, $\sim 10^{41}$ s$^{-1}$, implying low photoevaporation rates, $\sim 10^{-10} \ \mpy$, almost two orders of magnitude lower than our calculated FUV mass loss rates. 

We end with a caveat that the disk evolution scenarios and derived disk lifetimes  in this preliminary analysis are still somewhat qualitative in nature. Our results depend on the analytical calculations of Adams et al.(2004) for the evaluation of the photoevaporative flows, which are launched subsonically and travel significant distances before passing through the sonic point. Although the subsonic flow analysis is supported by hydrodynamical models (see Font et al. 2004 for flow inside of $r_g$), there are uncertainties inherent in the assumptions, especially that of the isothermality of the flow. It is conceivable that gas temperatures may rise due to heating at lower densities along a flow streamline and increase the derived mass loss rates. Magnetic fields in the disk will provide additional support against gravity and make it easier for flows to launch from the disk. On the other hand, the flow might cool as it escapes and  we may have therefore overestimated photoevaporation rates. The resolution of these issues requires hydrodynamical models of the flow and a solution of flow energetics. We will address these issues in future studies.

\subsection{What is the significance of X-rays for photoevaporation?}
Young stars have high X-ray luminosities which can result in significant disk heating. X-rays can therefore potentially cause substantial photoevaporative mass loss. Our earlier study (GH09) found that X-rays do not induce significant photoevaporation by themselves,  but can enhance the mass loss caused by FUV photons through their ionization of gas. Ercolano et al. (2008, 2009), however, find that X-ray heating  may result in substantial mass loss. Their estimated photoevaporation rates ($\dot{M}_{pe} \sim 10^{-9}\mpy$) are lower in their later, improved models (ECD09) than their earlier work (Ercolano et al. 2008), but nevertheless still substantially higher than what GH09 obtain for their pure X-ray heated disk model.  We note that the two works differ considerably in their treatment of X-ray heating and ionization. We have attempted to make a careful comparison of the different assumptions and physics to resolve the discrepancies, using results kindly provided to us by Barbara Ercolano.  

We find that the disparity in photoevaporation rates between GH09 and ECD09  can be attributed to two main differences (i) the hard X-ray spectrum adopted by GH09 (and the standard model of this paper) versus the soft spectrum of ECD09 and (ii) the lack of molecular coolants in ECD09.  The static, thermo-chemical disk models indicate an almost negligible photoevaporation rate for a disk heated by X-rays alone (GH09).  For a hard X-ray spectrum peaking at $\sim2$ keV as we assume, most of the X-ray energy is absorbed at  gas column densities to the star  $\sim 10^{22}$ cm$^{-2}$, at deeper layers in the disk. In the absence of FUV irradiation, there is very little photodissociation of molecules at these layers. The increased abundances of strong coolants such as CO and H$_2$O result in lower gas temperatures, and therefore low photoevaporation rates. Furthermore, the cooler disk surface results in less flaring of the the disk, in turn intercepting a lower fraction of the X-ray photon flux from the star. Low ionization fractions of the gas also keep the X-ray heating efficiency relatively low ($\sim$10-20\% for atomic gas and $\sim$40\% for molecular gas).   ECD09 assume an X-ray spectrum  with significant fluxes at $\sim 0.1-0.2$keV, and these soft X-ray photons are absorbed at  much lower column densities to the star, and high up at the disk surface. They  ignore molecules in their disk models, and assume that the gas is atomic in 
their X-ray layer. With less cooling and greater heating rates due to the soft spectrum,  they obtain photoevaporation rates $\sim 10^{-9} \mpy$ for their standard disk model.  Considering that soft X-rays  may play an important role in disk photoevaporation (ECD09) and also that soft X-ray excesses are often observed in young, accreting stars (G\"udel \& Telleschi 2007, Preibisch 2007), we solve for the time evolution of a disk using a soft X-ray spectrum similar to that adopted by ECD09 ($L_X(E)\propto E^{-1}$ for $E<2$keV). We do, however, keep the total X-ray luminosity the same as in our standard (hard) X-ray spectrum. 

\begin{figure}
\plotone{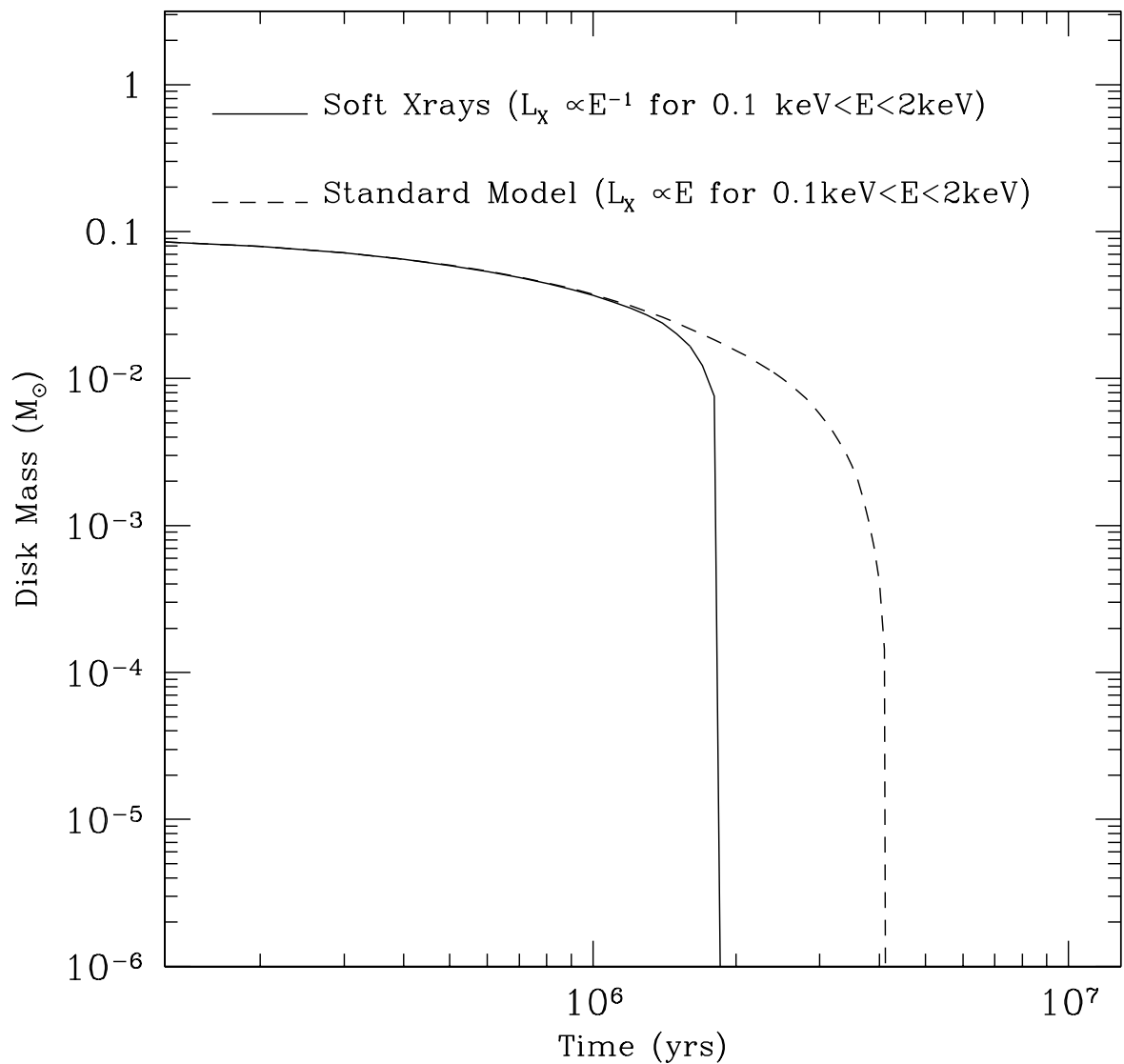}
\caption{The disk mass with time for a central $1M_{\odot}$ star with a soft X-ray spectrum is shown (solid line). Both models have a total X-ray luminosity of \pow{2}{30} erg s$^{-1}$. The disk is also subject to FUV and EUV photoevaporation. The standard model with a hard X-ray spectrum is also shown for comparison(dashed line). Soft X-rays can drive substantial mass loss in the disk and are found to shorten disk lifetimes by a factor of $\sim 2$. }
\label{fig9}
\end{figure}

Figure~\ref{fig9} shows how the mass of the disk decreases due to photoevaporation and accretion onto the central star with a soft X-ray spectrum, FUV and EUV.  The standard model disk evolution (X-rays peaking at 1keV) is also shown for comparison. The disk mass evolves as in the standard model for almost $1.5\times 10^6$ years, after which it suddenly declines. This is because at initial times, the accretion rates are high, and the disk wind column densities too large for penetration by soft X-rays.  The disk evolves under the actions of viscosity and FUV and hard X-ray photoevaporation as in our standard model (dashed line). The mass of the disk declines until FUV and hard X-ray photons begin to deplete the gas at $r\approx 5$ AU, just prior to gap opening
at $1.5\times 10^6$ years.  Disk wind column densities need to be lower than $\sim 10^{20}$ cm$^{-2}$ ($\dot{M}_w \sim 10^{-9}\  \mpy, \dot{M}_{acc} \sim 10^{-8} \ \mpy$) before the $\sim 0.1-0.2$ keV photons can penetrate the wind and heat the disk surface. The soft X-rays which were attenuated prior to this epoch can now irradiate the inner rim 
as the inner disk is depleted by viscosity and accretion rates onto the star drop.  
Soft X-ray photons are quite efficient in heating the gas and cause vigorous photoevaporative mass loss once they are incident on the disk.  The entire disk dissipates very rapidly from here on, with the disk surviving in total for $\sim 2\times 10^6$ years. The disk lifetime decreases by a factor of $\sim 2$ when adopting this soft X-ray spectrum. However if the soft X-ray component depends on ongoing accretion, it is not certain that soft X-rays would play a significant role in rapid disk dispersal after the formation of an inner hole by FUV photoevaporation. 

Clearly, further investigations of the importance of X-rays for disk photoevaporation are needed. Observations indicate that the nature of the X-ray spectrum changes with stellar accretion, and this is an important issue with the current models. Non-accreting weak line T Tauri stars do not show soft excesses. If the soft excess is indeed due to accretion flows cooling the hot plasma in the coronal  X-ray emitting regions (G\"udel \& Telleschi 2007), then once accretion halts, the spectrum would become harder. 
A simple approach to modeling the time dependence of the X-ray spectrum, similar to that available for the FUV spectrum (Calvet \& Gullbring 1998) is needed. In the future, we will solve for the full chemistry and thermal balance in the disk using an evolving X-ray spectrum and study disk evolution due to photoevaporation by EUV photons and a time-dependent  FUV luminosity and X-ray spectrum.

\subsection{Effects of changes in the viscosity parameter $\alpha$}
There is considerable uncertainty in the value of the parameter $\alpha$ used for determining the disk viscosity. Numerical models of the MRI in disks and indirect inferences from observational studies point to values that range from $\sim 10^{-1}-
10^{-4}$ for a typical disk around a 1 \ms\ star (e.g., King et al. 2007, Andrews et al. 2009).  In the earlier sections, we have shown that both viscosity and FUV-induced photoevaporation are  important in driving disk evolution. Both of these effects in turn depend on the poorly constrained viscosity parameter $\alpha$, for which we have thus far adopted a canonical value of $0.01$ for a 1\ms\ star. The accretion rate in the disk is proportional to $\alpha$, and the accretion-generated component of the FUV luminosity is derived from the accretion rate onto the star. We would therefore expect that disk lifetimes are affected by the choice of $\alpha$. 
 In order to study the effects of varying $\alpha$ on disk evolution, we consider models of disks around 1\ms\ stars with three different values, $\alpha=0.001, 0.01$ and $0.1$, assumed to be constant with disk radius for simplicity.

\begin{figure}
\plotone{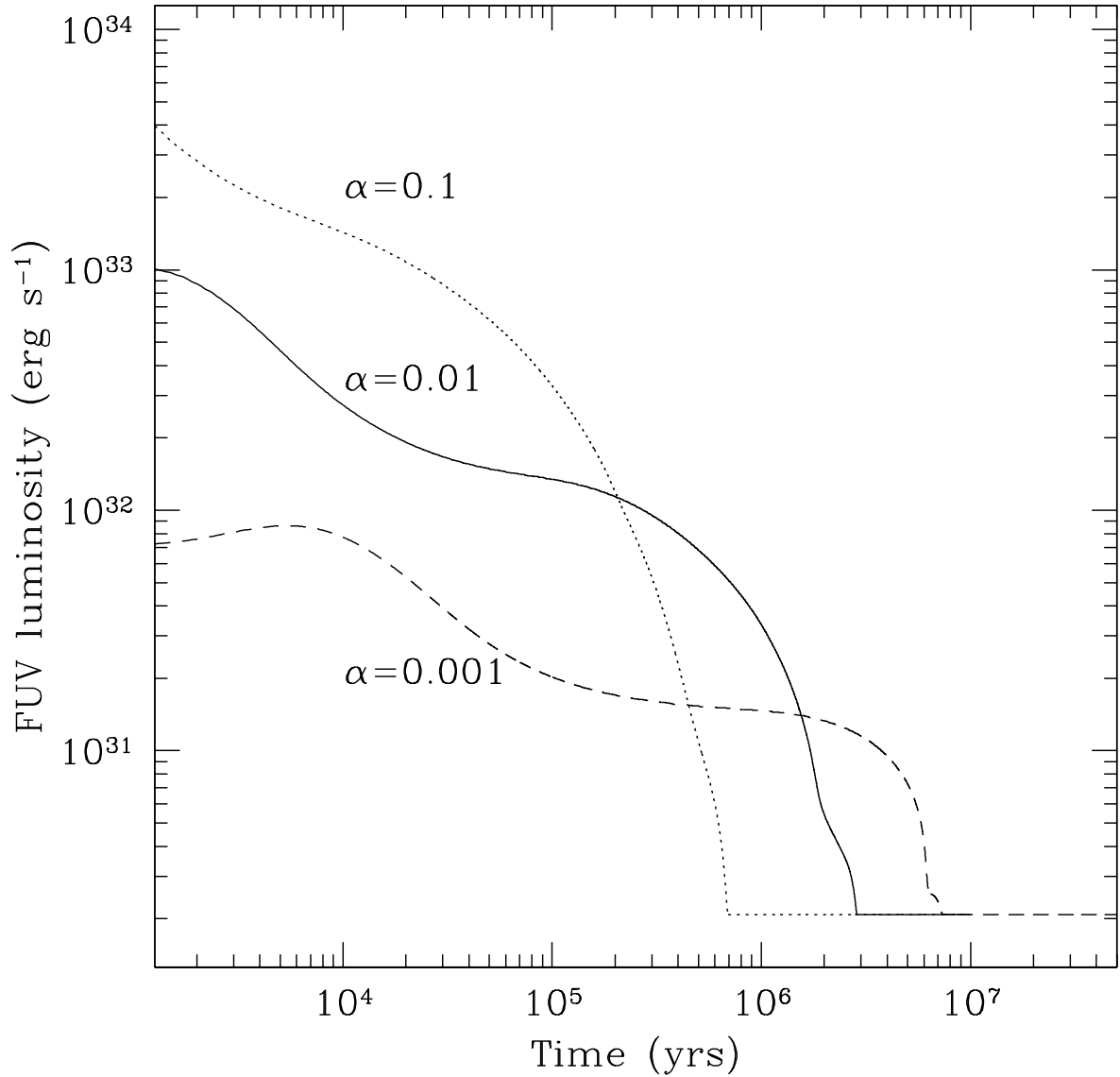}
\caption{The FUV luminosity is shown as a function of time for models of disks around 1\ms\ stars for three different values of the viscosity parameter $\alpha$. Higher values of $\alpha$ result in higher accretion rates onto the star and higher accretion-induced FUV luminosities. Whan accretion ceases, all three curves eventually plateau at the same chromospheric FUV luminosity. }
\label{fig10}
\end{figure}

\begin{figure}
\plotone{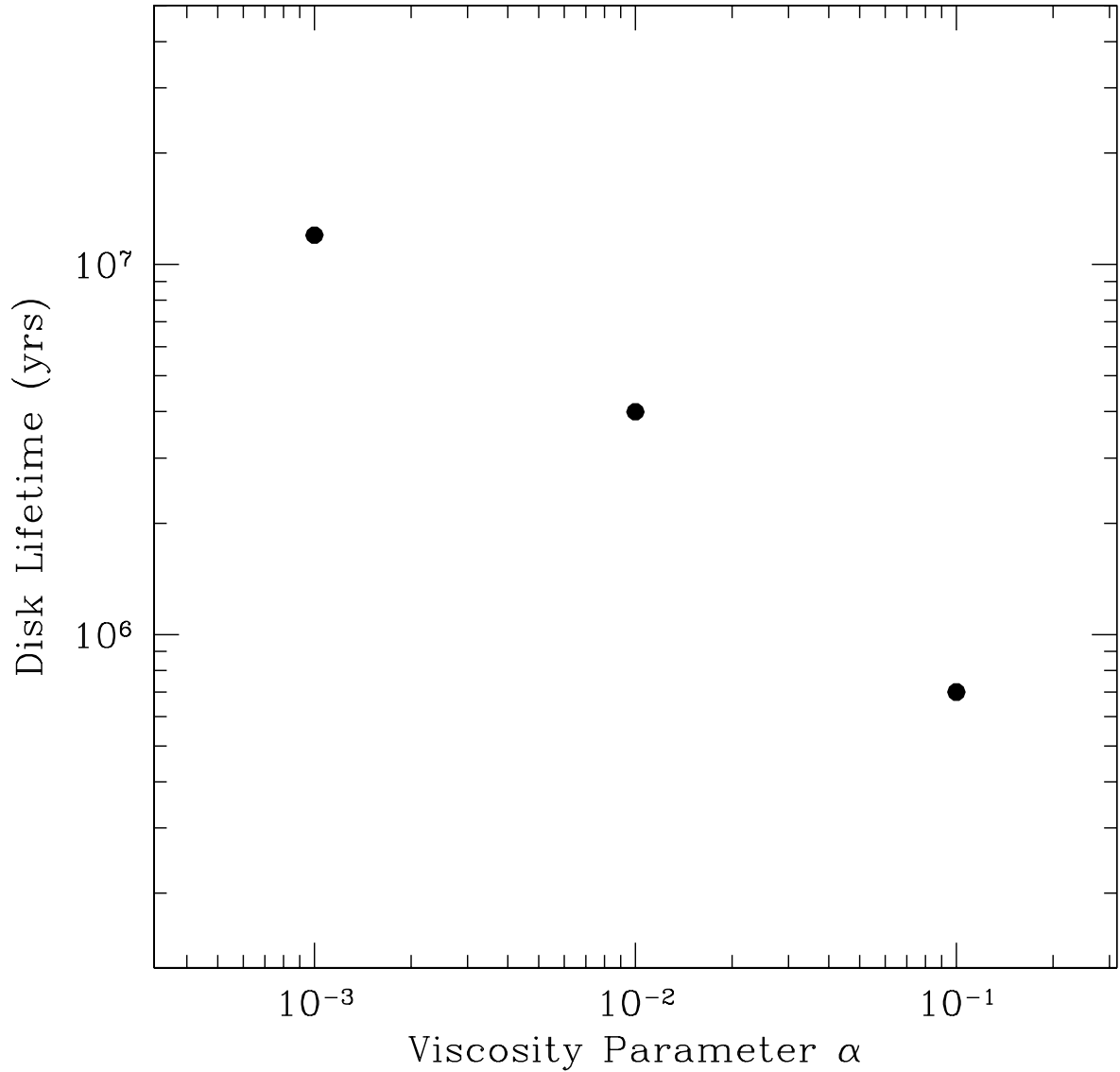}
\caption{ Disk lifetimes around a 1\ms\ star, defined by the complete dispersal of the disk, are shown for three different value of the viscosity parameter, $\alpha$.  The dispersive effects of viscosity and FUV-driven photoevaporation are  both enhanced for higher $\alpha$-values and disks lifetimes are shorter.  Disks with low $\alpha$, on the contrary, survive longer.}
\label{fig11}
\end{figure}

We find that lower values of $\alpha$ result in longer disk lifetimes and that higher values cause the disk to disperse sooner. 
Figure~\ref{fig10} shows the time-dependent FUV luminosity from the star for the three different disk models. When $\alpha$ increases, the higher viscosity results in a higher accretion rate onto the star and a higher accretion-generated FUV flux from the star. The 
dotted line shows the FUV luminosity for a model with $\alpha=0.1$, which is higher than that for our standard model ($\alpha=0.01$) at early stages of evolution. However, the higher rates of FUV photoevaporative mass loss cause the disk to lose mass faster, resulting in a decreasing surface density which drives down $\dot{M}_{acc} (\propto \alpha  \Sigma)$ and $L_{FUV}^{acc}$. The decrease in $L_{FUV}^{acc}$ in turn decreases photoevaporation rates and to some extent, disk evolution is self-regulated.  The FUV luminosity at later times drops rapidly below that of the standard model. The creation of a gap and an inner disk hole subsequently halts accretion and the FUV luminosity then levels off at  the chromospheric value. Similarly, the disk with low $\alpha$ of $0.001$ evolves slower than the standard disk and shows an $L_{FUV}^{acc}$ that is initially lower than the standard model (dashed line).  Figure~\ref{fig11} shows the disk lifetimes for the three different models. The dependence of the disk lifetime on $\alpha$ is seen to shallower than  might be expected, mainly due to  the complex nature of the feedback between the accretion rate and FUV photoevaporation rate as described above.  Gaps (and inner holes) are formed in all three disk models.  For the values of $\alpha$ considered here $0.001,0.01$ and $0.1$, the disk lifetimes are \pow{1.2}{7}, \pow{4}{6} and \pow{7}{5} years respectively. For low $\alpha$ and low accretion rates, gaps form earlier in evolution and a smaller fraction of the disk mass is lost due to accretion versus photoevaporation. Disk survival times after the creation of a hole are $\sim$ 40\%, 30\% and 10\% of their total lifetimes and the fraction of the disk mass lost due to photoevaporation is 0.68, 0.48 and 0.35 for $\alpha=0.001,0.01$ and $0.1$ respectively.

\subsection{Disk lifetimes and stellar mass}
Photoevaporation rates critically depend on the stellar radiation field which is a strong function of the stellar mass, and hence disk lifetimes may also depend on the mass of the central star. The bolometric luminosities of stars increase considerably as stellar mass increases, and for $M_*\gtrsim 3 {\rm M}_{\odot}$,  the corresponding X-ray and FUV luminosities are also high.  Furthermore, early-type (OB) massive stars are hotter and their photospheric radiation peaks at ultraviolet wavelengths.  They, therefore, have large intrinsic photospheric UV fluxes. Photoevaporation rates may as a result be expected to be higher for higher mass stars. On the other hand, disks around more massive stars are also more massive  and in a stronger gravitational field. Therefore  the disk has a larger mass reservoir and the gas is more bound,  which may work to counter the stronger radiation fields and increase disk lifetimes by photoevaporation.  Intermediate mass stars,  $M_*\sim 3 {\rm M}_{\odot}$, lack strong photospheric FUV components and are also chromospherically less active as the stellar interiors are not convective. Their disks are relatively massive and in a strong gravitational field. These disks may be expected to be long-lived. 
Low-mass stars, $M_*\lesssim 3 {\rm M}_{\odot}$, have less massive disks which are in a weaker gravitational potential, but relatively strong accretion-generated radiation fields. Chromospheric activity in low mass stars also results in higher X-ray luminosities relative to  mass than intermediate-mass stars. All these may combine to result in short disk lifetimes. 
There might thus be a peak in disk lifetimes for intermediate-mass stars. 
 Longer disk lifetimes may create conditions favorable for the formation of planetary systems, and facilitate the formation of gas giant planets.  A peak in disk lifetimes with stellar mass may imply a characteristic stellar mass for the likelihood of planetary system formation, a potentially interesting result with many consequences.  We attempted to derive this dependence in our earlier static models (GH09) and found that for stars with  $M_*\lesssim 3{\rm M}_{\odot}$, disk lifetimes were fairly constant, whereas more massive stars destroy their disks rapidly. We address this question again with our new time-dependent analysis.

\begin{figure}
\plotone{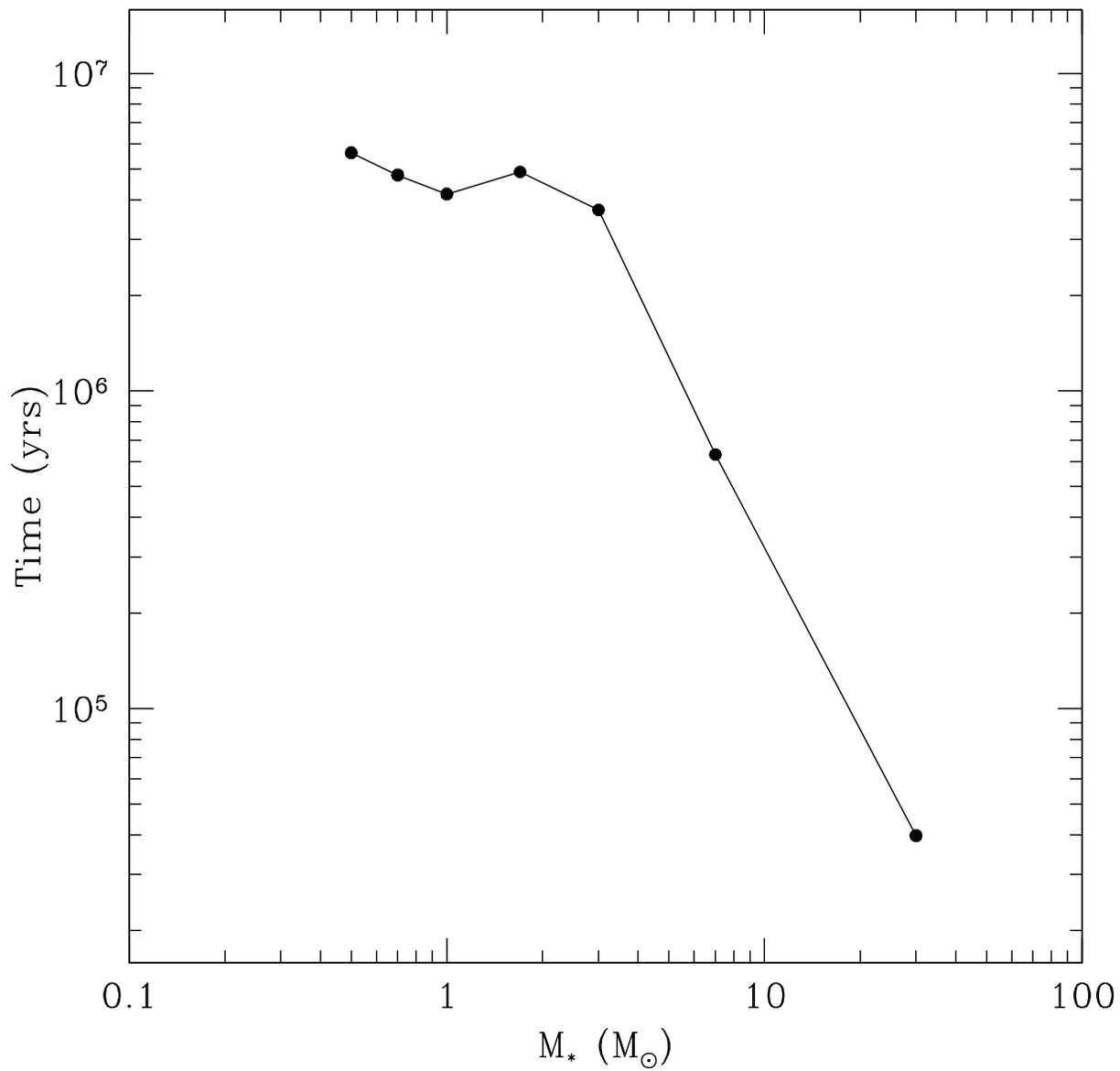}
\caption{Disk lifetime as a function of central star mass. Disks around stars with $M_*\lesssim 3{\rm M}_{\odot}$ survive for \app \pow{3-6}{6} years, while disks around higher mass stars are short-lived. The initial disk masses are $0.1M_*$,  EUV and X-ray luminosities are in Table 1, and the other disk parameters are listed in Table 2. The FUV luminosity varies with accretion rate as discussed in the text. 
 }
\label{fig12}
\end{figure}

We investigate the dependence of disk lifetimes on stellar mass by  considering several models where we vary the central star mass from 0.5\ms\ to 30\ms.  Figure~\ref{fig12} shows model disk lifetimes for different stellar masses. For  $M_*\gtrsim 3{\rm M}_{\odot}$, the disks survive for less than $10^6$ years and the lifetimes are shorter for more massive stars. Strong radiation fields rapidly photoevaporate these disks.  The presence of any mass dependence for  $M_*\lesssim 3{\rm M}_{\odot}$ is inconclusive. There appears to be a slight trend of disk lifetimes increasing as stellar mass decreases for  $M_*\lesssim 1.7 {\rm M}_{\odot}$, but given the uncertainties inherent in the modeling, this may not be significant. Typical disk lifetimes for stars with  $M_*\lesssim 3{\rm M}_{\odot}$ are $\sim 3-6 \times 10^6$ years. 

\section{Summary}
The short observed lifetimes of protoplanetary disks (e.g. Haisch et al. 2001) entail a mechanism for their dispersal. While  a combination of EUV irradiation and viscous evolution has been proposed to be responsible for disk destruction (e.g. Clarke et al. 2001, ACP06), we demonstrate that for typical disk masses and stellar EUV luminosities, this is insignificant. Using static thermo-chemical models, we had earlier proposed that photoevaporation caused by FUV/X-ray photons from the star is more effective in disk dispersal (GH09), and in this paper we follow this up with a time-dependent model that includes viscosity. 

Our new results confirm the importance of FUV and X-rays in disk dispersal. We use a 1D time-dependent model for surface density evolution due to viscosity and photoevaporation, combined with a 1+1D model for the dust and gas density and temperature structure. We use simplified approaches for obtaining the gas temperature, 
and eliminate the complexity of solving a chemical network.  We allow for the attenuation of the stellar FUV, EUV and X-ray flux by the disk wind that typically accompanies accretion (e.g., Shu et al. 1994, White \& Hillenbrand 2004). We study the evolution in surface density distribution of a viscous accretion disk as it is undergoes photoevaporation by EUV, FUV and X-ray photons from the central star.  

We find that the disk is first irradiated by FUV photons and high energy ($\gtrsim 1$keV) X-rays and at later, lower accretion rate epochs, by soft X-rays and EUV photons. 
The combination of FUV photoevaporation with viscous accretion onto the star and viscous spreading to the outer photoevaporation zones causes a decline of the surface density at all radii with time.   FUV photons can drive gaps in the disk, an important new result of this paper. We have used a time-dependent accretion-induced FUV luminosity to model the evolution of the disk, and find that FUV luminosities remain high enough at later epochs to drive gap formation. Another significant result is that after the inner hole is created, the remaining disk can survive for fairly long timescales, $\sim 10^6$ years. 
 EUV, at typical luminosities of $10^{-3}$L$_*$, are mostly insignificant for disk evolution, both in terms of removing disk mass and forming gaps in the disk.   We find the disk lifetime for a $0.1$M$_{\odot}$ around a solar-mass star is $\sim 4\times 10^6$ years. Soft X-rays, if they persist at low-accretion rate epochs,  can influence disk photoevaporation and reduce disk lifetimes by a factor of $\sim 2$.   Disk lifetimes are also inversely proportional to the assumed value of the viscosity parameter, $\alpha$. Disk lifetimes are found to be nearly independent of central star mass for  $M_*\lesssim 3{\rm M}_{\odot}$, and $\sim$ a few $10^6$ years.  Disks around more massive stars are short-lived, and are $\lesssim 10^5$ years
for $M_* \gtrsim 10 {\rm M}_{\odot}$. 

From our calculated gap formation timescale and total disk lifetime, the expected  fraction of disks with inner holes, or transition disks,  would be $\sim$ 25\%, whereas the observed fraction is believed to be significantly lower. We find that gaps form in all our FUV/X-ray photoevaporation models. The transition timescale has been previously estimated as $\sim 10^5$ years, based on statistics of primordial disks in young, 3Myr-old clusters (Skrutskie et al. 1990; Hartigan et al. 1990; Simon and Prato 1995; Wolk and Walter 1996). More recent studies of older clusters (e.g. Haisch et al. 2001, Hernandez et al. 2007, Currie et al. 2009) indicate that the lifetime of primordial disks surrounding many young stars is $\sim 3-5 \times 10^6$ years, and coupled with estimates of total disk lifetimes suggest that  the transition disk phase may in fact be longer, $\sim 2-3$ Myrs (Currie et al. 2009). This conclusion is also supported by recent studies of the Coronet cluster, where the transition disk population is nearly the same as the primordial disk population (Sicilia-Aguilar et al. 2008), indicating similar durations for both phases.  Different transition disk lifetimes may perhaps reflect  differences in  the viscosity parameter $\alpha$, as we find that disks with low $\alpha$ generally spend a longer fraction of their life as transition disks, whereas higher $\alpha$ values indicate a shorter duration of the transition disk phase. 
Transition disks may however also result from processes other than photoevaporation, such as grain growth, planet-disk dynamics  or binarity (e.g., Najita et al. 2007). We note that transition disks caused by EUV photoevaporation tend to have very low mass accretion rates onto the central star, whereas Najita et al (2007) point out that they often have significant accretion rates suggesting that the inner disks are gas rich and dust poor. Hydrodynamic models of FUV/X-ray photoevaporation in disks with low dust opacity per H nucleus are needed to see how much of the photoevaporated rim material can accrete onto the central star. 

We conclude that photoevaporation by FUV and X-rays combined with viscous evolution  is a viable mechanism for destroying disks around young stars. The lifetimes derived here, using typical parameters for $\alpha, L_X$ and $L_{FUV}$, are in accordance with observed disk lifetimes (e.g., Haisch et al. 2001, Hillenbrand 2005).  Disk lifetimes appear to be long enough to allow the formation of giant planets and perhaps planetary systems around all stars with $M_*\lesssim 3 {\rm M}_{\odot}$.  Our calculated disk lifetimes closely correspond with observationally derived disk lifetimes, supporting the role of FUV/X-ray photoevaporation in disk destruction. 

\acknowledgements
We thank Barbara Ercolano and Richard Alexander for helpful discussion during the course of this work and for generously providing data for comparisons. We thank the referee for a very careful and thorough reading of the manuscript and many useful comments that improved the paper.  Gorti and Hollenbach acknowledge financial support by research grants  from  the NASA Astrophysics Theory Program (ATP04-0054-0083), the NASA Astrobiology Institute and NSF(AST0606831).

\begin{table}
\centering
\caption{Standard Stellar Input Parameters as a Function of Mass}
\label{table1}
\begin{tabular}{cccccc}
\\
\tableline
\tableline
M$_*$ & R$_*$ & Log T$_{eff}$ & Log L$_*$  & Log L$_X$ 
& Log $\phi_{\rm EUV}$ \\
(M$_{\odot}$) & (R$_{\odot}$) & (K) & (L$_{\odot}$) &(erg s$^{-1}$)  & (s$^{-1}$)  \\
\tableline
0.3 & 2.30 & 3.52 & -0.26 & 29.6 & 39.9    \\
0.5 & 2.12 & 3.57 & -0.03  & 29.8 & 40.1 \\
0.7 & 2.54 & 3.60 & 0.24  & 30.2 & 40.5  \\
1.0 & 2.61 & 3.63 & 0.37   & 30.4 & 40.7 \\
1.7 & 3.30 & 3.66 & 0.70  & 30.7 & 41.0 \\
3.0 &4.83 & 3.70 & 1.17 &   28.7& 39.0 \\
7.0 & 3.22& 4.31 & 3.23 & 30.8& 44.1\\
30.0 &  9.25 & 4.54 & 5.69& 33.3& 48.7\\
\tableline
\end{tabular}
\end{table}

\begin{table}
\centering
\caption{Fiducial Disk Model - Input Parameters}
\label{fid}
\begin{tabular}{ll}
\\
\tableline
\tableline
Disk mass & 0.1 M$_*$ \\
Initial Surface density & $\Sigma(r) \propto r^{-1}$  \\
Viscosity Parameter $\alpha$ & $0.01 \times (M_*/M_{\odot}$) \\
Inner disk radius & 0.1 AU \\
Outer disk radius & 200 AU \\
Gas/Dust Mass Ratio & 100 \\
Dust grain size  & $0.3 \mu $m  \\
\tableline
\end{tabular}
\end{table}

\end{document}